\documentclass[a4paper,fleqn,preprint]{cas-dc}

\usepackage[numbers,sort&compress]{natbib}

\def\tsc#1{\csdef{#1}{\textsc{\lowercase{#1}}\xspace}}
\tsc{WGM}
\tsc{QE}

\usepackage[english]{babel}
\usepackage[utf8]{inputenc}
\usepackage[T1]{fontenc}
\usepackage{tgtermes}
\usepackage{cleveref}
\crefname{subsection}{subsection}{subsections}
\usepackage{todonotes}
\usepackage{placeins}

\usepackage{multirow}
\usepackage{siunitx}
\usepackage{graphicx}
\usepackage{dcolumn}

\DeclareSIUnit\angstrom{\text {Å}}

\usepackage[caption=false]{subfig}
\usepackage{bm}

\DeclareSIUnit{\atmospheric}{atm}

\begin{document}
\let\WriteBookmarks\relax
\def\floatpagepagefraction{1}
\def\textpagefraction{.001}

\shorttitle{Characterising anisotropic diffusion of small molecules and the effect of local drift in confinement}    

\shortauthors{H\"ollring et al.}  

\title [mode = title]{Anisotropic molecular diffusion in confinement I: \\ Transport of small particles in potential and density gradients}

\author[1]{Kevin H\"ollring}[orcid=0000-0002-9497-3254]
\credit{Conceptualization, Methodology,  Software, Formal analysis, Investigation, Writing - Original Draft, Visualization, Data Curation}

\author[1]{Andreas Baer}[orcid=0000-0001-7943-8846]
\credit{Investigation, Simulation, Writing - Original Draft (methods)}

\author[2,1]{Nataša Vučemilović-Alagić}[orcid=0000-0001-5841-7181]
\credit{Investigation, Design and Execution of Simulations, Writing - Original Draft (methods)}

\author[2]{David M. Smith}[orcid=0000-0002-5578-2551]
\credit{Supervision of the design of N.V.A.'s simulation}

\author[1,2]{Ana-Sunčana Smith}[orcid=0000-0002-0835-0086]
\credit{Conceptualization, Writing - Review \& Editing, Project coordination, Funding acquisition, Resources, Data Curation}
\ead{ana-suncana.smith@fau.de, asmith@irb.hr}

\cormark[1]
\fnmark[1]

\affiliation[1]{organization={PULS Group, Institute for Theoretical Physics, FAU Erlangen-N\"urnberg},
addressline={Cauerstra\ss{}e 3},
postcode={91058}, 
city={Erlangen},
country={Germany}}

\affiliation[2]{organization={Group of Computational Life Sciences, Department of Physical Chemistry, Ru\dj{}er Bo\v{s}kovi\'{c} Institute},
addressline={Bijeni\v{c}ka 54},
city={Zagreb},
postcode={10000},
country={Croatia}}

\cortext[1]{Corresponding author}
\fntext[1]{Tel: +49 91318570565; Fax: +49 91318520860}

\date{\today}

\begin{abstract}
\indent
\paragraph{Hypothesis:}
Diffusion in confinement is an important fundamental problem with significant implications for applications of supported liquid phases. 
However, resolving the spatially dependent diffusion coefficient, parallel and perpendicular to interfaces, has been a standing issue.
In the vicinity of interfaces, density fluctuations as a consequence of layering locally impose statistical drift, which impedes the analysis of spatially dependent diffusion coefficients even further. 
We hypothesise, that we can derive a model to spatially resolve interface-perpendicular diffusion coefficients based on local lifetime statistics with an extension to explicitly account for the effect of local drift using the Smoluchowski equation, that allows us to resolve anisotropic and spatially dependent diffusivity landscapes at interfaces.
\paragraph{Methods and simulations:}
An analytic relation between local crossing times in system slices and diffusivity as well as an explicit term for calculating drift-induced systematic errors is presented.
The method is validated on Molecular Dynamics simulations of bulk water and applied to simulations of water in slit pores.
\paragraph{Findings:}
 After validation on bulk liquids, we clearly demonstrate the anisotropic nature of diffusion coefficients at interfaces.
 Significant spatial variations in the diffusivities correlate with interface-induced structuring but cannot be solely attributed to the drift induced by local density fluctuations.
\end{abstract}

\begin{graphicalabstract}
\includegraphics{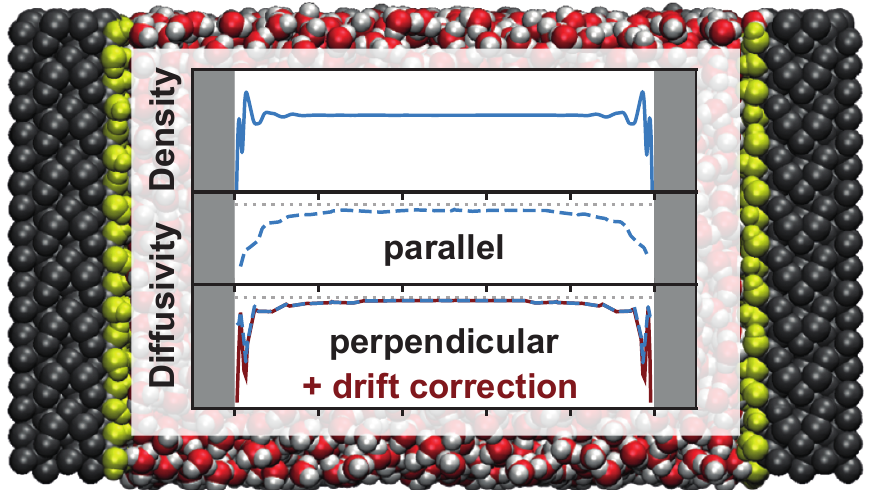}
\end{graphicalabstract}

\begin{keywords} 
transport coefficient \sep diffusion in pores \sep porous materials \sep anisotropic diffusion \sep diffusion at interfaces \sep drift and density gradient
\end{keywords}

\maketitle

\section{Introduction}
Transport in strongly confined geometries, such as in porous materials or thin films, is a fundamental problem in physics with direct applications in chemical sciences, engineering, biophysics and geosciences \cite{kramers1940brownian,marion2021water, lee2017real,siretanu2014direct}.
Most generally, the presence of interfaces breaks the symmetry of the system, impeding orthogonal molecular motions \cite{kestin1988transport}. 
Already on the hydrodynamic level, this poses challenges for theoretical modelling and experimental exploration, while resolving molecular details adds a strong multi-scale component into the problem \cite{lhermerout2018ionic,gebbie2017long}. 
Close to solid interfaces, the interactions between the liquid and the solid phase cause layering effects \cite{tournassat2016molecular,somers2013review,perez2017scaling}. 
This can both hinder or promote diffusive transport depending on the specific properties of the materials involved, and the direction of movement \cite{han2006brownian,pande2015forces}. 
The consequence is anisotropic mobility parallel and perpendicular to the confining surface \cite{mittal2008layering,fernandez2004self,nordanger2022anisotropic}.

It is typically difficult to account for the molecular nature of transport in confinement using analytic theory approaches \cite{smith2016electrostatic}.  
Therefore the modelling method of choice are molecular dynamics simulations, where the molecular details can be fully sampled while the long scale dynamics can be accessed with sufficient computing power \cite{fedorov2008towards,merlet2014electric,salanne2011polarization,kondrat2014accelerating,salanne2012including,salanne2015simulations,padua2007molecular,canongia2006nanostructural,canongia2004modeling,canongia2004molecular, horstmann2022structural}. 
However, in this case, the transport coefficients need to be extracted from recorded trajectories.

Several techniques have been established for that purpose. 
The most broadly used method relies on the Green--Kubo formalism, which employs velocity auto-correlation functions (VACF) to derive diffusion coefficients \cite{kubo1957statistical,green1954markoff}.  
While easy to apply, this approach is reliant on appropriate simulation procedures to produce the required correlation observations \cite{zhou1996green,fernandez2004self,fong2021ion,zwanzig1970hydrodynamic}.  
Equally common is the Einstein approach \cite{micheletti2008optimal,vella2019fick,fong2021ion,sicard2021position}, which derives diffusion coefficients from (positional) mean square displacement (MSD)
\footnote{\textbf{Abbreviations:} Mean Square Displacement (MSD), 
velocity auto-correlation functions (VACF),
Simple Particle Model (SPM), 
Simple Particle Model with drift (SPM+d), 
Solid-Liquid (SL), 
Liquid-Vacuum (LV), 
Solid-Liquid-Solid (SLS), 
Interface Normal Number Density (INND), 
Partial Differential Equation (PDE)}
 \cite{uhlenbeck1930theory}.
Both of these methods have been adapted to address specific confinements, represented by reflecting boundary conditions for point like objects diffusing with a spatially independent transport coefficient \cite{turkcan2012bayesian,nagai2020position}. 
Using MSD and VACF is appropriate for the analysis of diffusive transport in the direction parallel to the interface, in layers that are sufficiently thin such that the necessary conditions concerning symmetry, isotropy, and homogeneity apply.
They are, however, not well-suited for the analysis along coordinates where the diffusivity is variable and affected by the confinement, i.e. the diffusion coefficient perpendicular to an interface. 
Still they have been applied to such scenarios with varying degrees of success \cite{holmboe2014molecular,hinczewski2010diffusivity}.

Dividing systems into slabs or layers to resolve spatial variability, comes, however, at a cost for the methods based on the MSD and the VACF, and results in a clear resolution limit.  
This limit is established by the fact that purely diffusive motion only sets in on the middle- to long-term timescale and that the sampling of sufficiently long trajectories is biased by the finite width of the layer in the orthogonal direction \cite{shalchi2011applicability,comer2013calculating}. Furthermore, convergence issues may appear \cite{best2010coordinate} which may even be severe \cite{bourg2011molecular}.

In recent times, a third family of methods has been used more frequently \cite{buchete2008coarse,bourg2012molecular,rosta2015free}. 
This class of models involves Markov-State-Model \cite{husic2018markov} and Bayesian approaches built from the ground up. 
In these approaches the space is systematically split into subspaces (slabs/slices) for particle positions \cite{renkin1954filtration,bourg2012molecular,buchete2008coarse}. 
The typical observable are transition rates \cite{hummer2005position,rohrdanz2011determination} or transition times \cite{perez2013slowparameters} between these subspaces that are linked to the underlying model parameters, such as diffusivity, via likelihood estimators assumed to reasonably model the analysed configuration. 
In the Bayesian approach specifically, the likelihood estimators are employed to derive a probability distribution on the parameters space to identify the most likely set of parameters underlying the observed time evolution. 
Through this approach, Markov State Models have been shown to perform as good as MSD/Green--Kubo approaches in unconfined geometries and surpassing their accuracy in confinement \cite{voisinne2010quantifying,comer2013calculating}.

The Markov State and Bayesian approaches, however, also suffer from certain constraints. 
They require an appropriate likelihood-estimator \cite{voisinne2010quantifying}, which may only be derived as an approximation and is not universally available. 
These methods also rely on a “good enough” a priori estimate of reasonable parameters \cite{comer2013calculating} to secure an accurate posteriori distribution. 
Furthermore, basing the analysis on transitions between the states yields relative behaviour, which may require calibration to an established baseline for the investigated liquid, instead of an absolute, purely local result. 
This problem is particularly evident in the jump-diffusion model \cite{bourg2011molecular}, which represents a subclass of  Markov-State-Models.  

The jump-diffusion model attempts to provide a link between the time spent in certain compartments of the system, the size of these compartments and average local diffusion coefficients \cite{bicout1998electron}. 
However, our own investigation of this relation showed vast discrepancies in absolute diffusivities close to a pore wall \cite{vucemilovic2019insights}.  
Even in the original derivation, the authors used it only to provide a qualitative and relative estimate of the evolution of the transport of a simple particle close to an interface. 
Furthermore, in its current formulation, the jump-diffusion approach does not account for statistical drifts resulting from the potential of mean force between the diffusing particle and the confining walls. 
The impact of such a drift is especially severe close to the interface, where the potential diverges, and where significant density variations of the solvent typically occur further impacting the effective potential of the diffusing particle. 
It is therefore not possible to evaluate the systematic error made by estimates close to interfaces, which is still a major challenge. 

In this work, we resolve these problems by expanding on the existing jump-diffusion approach. 
We first provide a precise formula linking the local diffusion coefficient of simple small particles to the observed mean duration of particle stays in a particular subspace, as a function of the subspace size. 
This enables us to provide absolute diffusivities without the need for a reference calibration. 
We furthermore perform a detailed analysis of the role of statistical drift by calculating the first-order correction to the basic drift-free model. 
Based on this description, we are able to analyse the anisotropic diffusion profile of water in a slit hydroxylated alumina pore and its coupling to the local density profile. 
As a result we clearly demonstrate the oscillatory behaviour of diffusive transport coefficients at relatively large distances from the pore wall. 
Interestingly, we find that the drift due to effective interactions of the water with the wall affects the results only at the contact with the wall and hence, the basic model is sufficient for quantitatively describing the behaviour throughout the centre region of the pore.  

\begin{figure}
\includegraphics[width=\columnwidth]{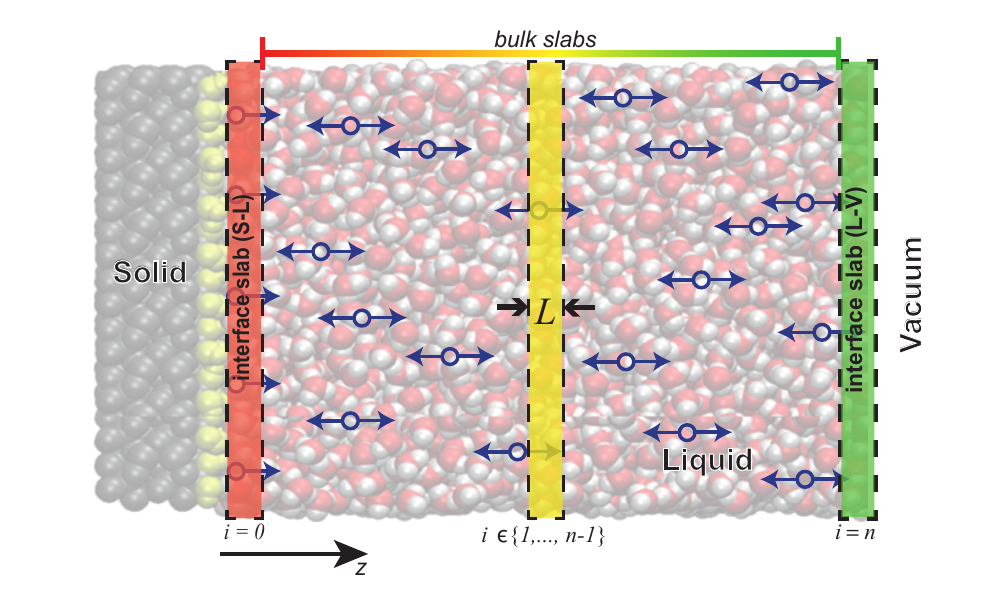}
\caption{\label{fig:scheme_system}\textbf{Scheme of a nanoconfined simulation box and the associated slicing:} 
The highlighted light red rectangle (left) represents a thin slab near a solid interface, light yellow (middle) corresponds to bulk-like slabs and light green (right) represents a thin slab near a vacuum interface. 
See text for details. \textit{(Figure adapted from \cite{vuvcemilovic2021computational})}}
\end{figure}

\section{Simple particle model for the perpendicular diffusion coefficients}
Our first goal is to determine the local diffusion coefficient $D_\perp(z)$ within a slice based on the life time of a simple, point-like particle within the slice. 
We first perform this calculation in the absence of any drift, or spatially dependent diffusivities within the subspace of interest. 
After the establishment of this basic link, we will discuss an extended model, accounting explicitly for the presence of non-zero drift. 
This allows for an estimate of the systematic error of the diffusion analysis as a consequence of neglecting drift in this so-called Simple particle model (SPM).

We base our analysis on a reduction of the liquid dynamics to movement along only one major axis, which we will refer to as the $z$-direction. 
The other dimensions are reduced under the assumptions of sufficient symmetry. 
In the presence of an interface, we assume $z$ to be interface-orthogonal (see e.g. \cref{fig:scheme_system}). 
This effectively 1D-system can then be cut into smaller slices of thickness $L$ (\cref{fig:scheme_system}). 
More complex extensions than a 1D-system are possible but beyond the scope of this work. 

We calculate the probability distribution $p(z,t)$ that a particle which is within the slice at time $t=0$ continuously remains within the slice until time $t\geq 0$ when it is found at position $z$ within the interval $z\in[z_i,z_i+L]$, where $z_i$ is the lowest $z$ coordinate of the interval. 
According to standing literature \cite{belousov2022statistical} $p(z,t)$ is  best described by the following Smoluchowski equation:
\begin{equation}
\partial_t p(z, t)= \partial_z ( D_\perp (z) \partial_z p(z, t)), z \in[z_i,{z_i+L}].\label{eq:fpe_raw_z}
\end{equation}
The solution for the average life time $\tau$ emerges from integrating the resulting probability distribution $p(t)$ that a random particle within the slice at time $t=0$ has not left the slice until time $t>0$ (see Appendix \ref{app:theory_spm}).
As the slice thickness $L$ is chosen by us and $\tau$ can be obtained from the analysis of MD trajectories, one can then use the established link to compute $D_\perp$.

With the assumption of constant particle density within the slab we can set the initial condition as $p(z,0)=const$. 
Now, the Smoluchowski equation can be solved for each slice independently, and there is no coupling over the boundary conditions between two neighbouring slabs. 
Integrating the distribution $p(z,t)$ then yields the prediction for the mean lifetime $\tau$, which is generally of the form
\begin{align}
\left\langle D_\perp  (z)\right\rangle  =const\times \frac{L^2}{\tau}.\label{eq:D_relation_basic_shape}
\end{align}
The constant prefactor is determined by the boundary conditions, which depend on the positioning of the slab relative to the interfaces or rather the type of slab we are investigating. 

\emph{Partitions within the fluid:}  
For bulk-like slabs, where the particles of the liquid can escape in both $z$ directions (index $\mathrm{B}$, i.e. yellow in \cref{fig:scheme_system}) we choose absorbing boundary conditions in both directions at $z=z_i$ and $z=z_i+L$. 
For our Simple Particle Model, this yields the following relation for $D_{\perp \mathrm{B}}$ (see appendix \ref{app:theory_spm} for detailed solution):
\begin{equation}
D_{\perp B }=\frac{1}{12}  \frac{L^2}{\tau_B}.\label{eq:D_perp_B}
\end{equation}

\emph{Interfacial slabs:} \Cref{eq:fpe_raw_z} is also solved in a scenario applicable to a slab at an impenetrable but otherwise non-interactable interface like a liquid-vacuum (LV) interface (index $\mathrm{LV}$, green in \cref{fig:scheme_system}).  
The slice boundary towards the vacuum is modelled to be reflecting, while the boundary towards the bulk liquid is treated as being absorbing. 
Accordingly, the particle is allowed to escape from the interface slice only to the next slice towards the bulk (see appendix \ref{app:theory_spm} for detailed solution methodology). 
Under these conditions the average diffusion coefficient $D_{\perp \mathrm{LV}}$ becomes:
\begin{equation}
D_{\perp \mathrm{LV}}=\frac{1}{3}  \frac{L^2}{\tau_\mathrm{LV}}.\label{eq:D_perp_LV}
\end{equation}

One can apply the same approach to a slab at the interface between the solid and the liquid (index $\mathrm{SL}$, i.e. red in \cref{fig:scheme_system}). 
However, the assumption of a vanishing external potential (and thus stochastic drift) may not be directly applicable in such a slice if there are strong molecular interactions (Coulomb forces or hydrogen bonds) \cite{atkin2007structure}.  
When the molecules of interest adsorb to the surface and are basically immobilised, as it happens for water close to hydrophilic interfaces or ionic liquids \cite{brkljaca2015complementary, vucemilovic2019structural}, the effective interface surface can be shifted beyond the adsorbed layer, and the derived result for an LV slab can be applied with good accuracy.

For the solution of \cref{eq:fpe_raw_z}, the spatial dependence of the diffusion coefficient $D_\perp (z)$ within the slab has been suppressed, and is replaced by its average value within the slice $\left\langle D_\perp\right\rangle= \left\langle D_\perp  (z)\right\rangle$ (i.e. $\partial_z D_\perp (z) \approx 0$). 
We additionally assumed a constant free energy background within each slab. 
These conditions  are  entirely  fulfilled  in  bulk liquids.  
In confined liquids, a non-constant statistical density profile of a particle of interest develops at the interfaces due to the effective interaction potential with the interface. 
At the extreme  points  of  that potential, the conditions of constant  background  potential and constant diffusion  are in essence correct. 
The assumption of constant potential is, however, only an approximation in between extrema. 

\subsection{Accounting for the influence of particle drift}
\begin{figure}
    \centering
    \includegraphics[width=.95\columnwidth]{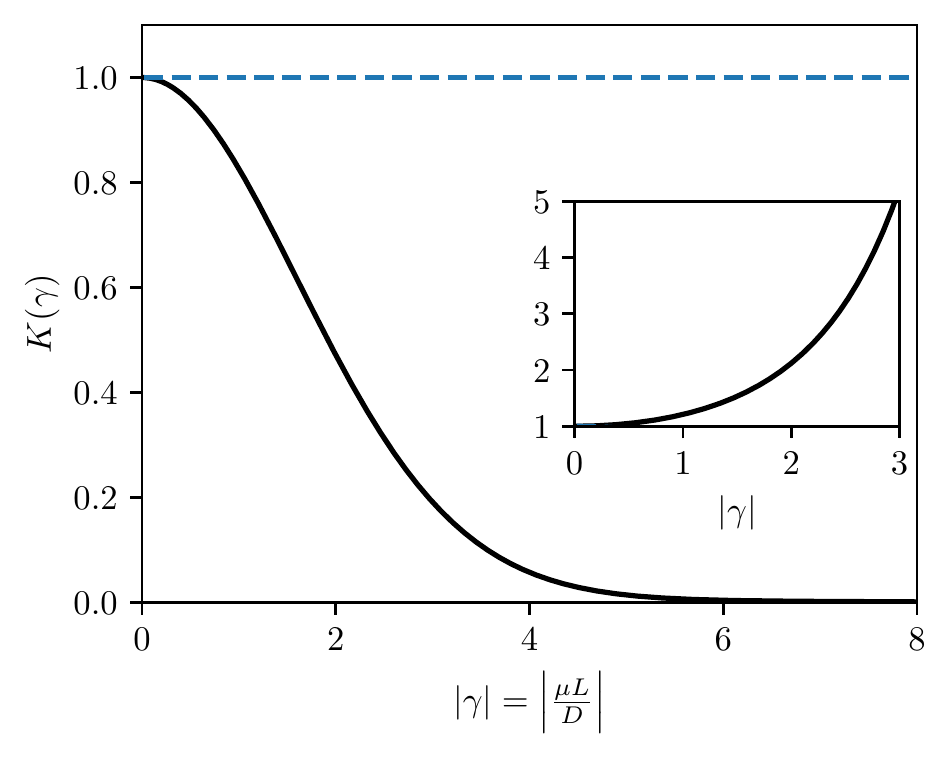}
    \caption{\textbf{Visualisation of the drift correction coefficient $K_\mathrm{B}(\gamma)$}. The plot shows the evolution of $K_\mathrm{B}(\gamma)$ (\cref{eq:drift_induced_correction}) with increasing drift magnitude from no correction (i.e. $1$) to the drift being the main contribution to lifetime (i.e. value $0$). The inset shows the relative overestimation of the resulting diffusion coefficient $D$ as a function of $\gamma$ when applying the pure SPM with $1$ representing the actual value. }
    \label{fig:drift_theory_coefficient}
\end{figure}

To address this aforementioned issue of the SPM, we now explicitly deal with the presence of a drift and analytically quantify the systematic error introduced by its omission.
Technically, we follow a similar approach as employed for the solution to the drift-free Smoluchowski equation. 
We derive the relation between $D$, $L$ and $\tau$ accounting for a constant drift $\mu$ induced by a linear change in the effective background potential across a single slice (see Appendix \ref{app:theory_spmd} for detailed derivation). 
Here, we will limit our analysis to the bulk-like slab as the most common subspace geometry. 
As a result, we arrive at a relation 
\begin{align}
    D_\mathrm{B}(L,\tau,\mu) =& \frac{1}{12} K_\mathrm{B}\left(\frac{\mu L}{D}\right) \frac{L^2}{\tau},
     \label{eq:equation_theory_spmd}
\end{align}
which we term the Simple Particle Model with a drift (SPM+d), with a correction factor 
\begin{align}
    K_\mathrm{B}\left(\gamma\right) =\frac{24}{\pi^4}\frac{\gamma^2}{ \cosh\left(\gamma\right)-1 } \sum_{n=1}^\infty \frac{ \left(1-(-1)^n \cosh\left(\frac{\gamma}{2}\right) \right)}{\left(n^2+\frac{\gamma^2}{4\pi^2} \right)^2}. \label{eq:drift_induced_correction}
\end{align}
The evolution of $K_\mathrm{B}$ is visualised in \cref{fig:drift_theory_coefficient}.
The interesting observation here is the square dependence of the correction on the relative drift amplitude $\gamma = \mu L /D$.  
This allows for the control of the systematic error via a reduction of $L$. 
Consequently, the SPM can then still be used in the presence of a gradient of the effective potential, when the slabs are sufficiently thin such that the change in density between the two boundaries is small compared to the average background, the latter being explicitly accounted for.
The systematic error introduced through the omission of the drift can then be expected to be reasonably small except for a very drastic potential changes e.g. immediately adjacent to an interface, where the SPM+d allows for a first-order correction.

\section{Validating the SPM model using bulk water}
\label{spm_epm_validation}
\label{water_bulk_investigation}

\begin{figure}[b]
    \centering
    \includegraphics[width=0.5\textwidth]{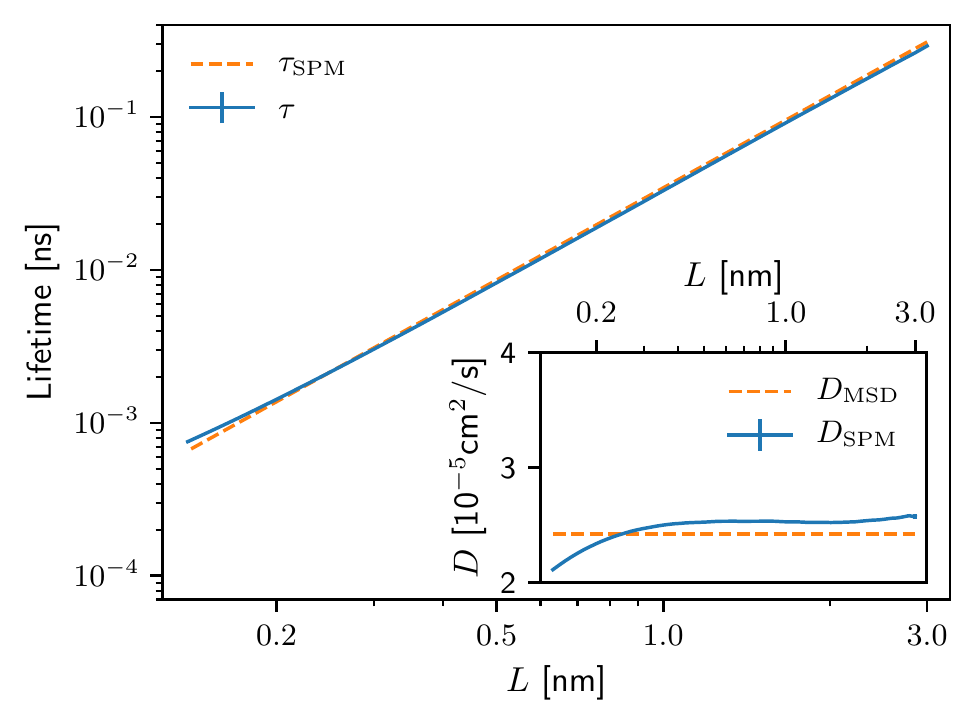}
    \caption{\textbf{Benchmark of the Simple Particle Model (SPM) on a bulk water system for various choices of slab thickness $L$:} 
    Great agreement of average observed lifetimes $\tau$ (blue, solid) with the prediction made by the SPM (orange, dashed) based on bulk MSD values for $D$.
    An exception is noticeable where the mean lifetime approaches the frame time difference of the simulation and the statistics thus overestimate the crossing time at small $L$. 
    (Inset) 
    Comparison of the value of $D$ obtained from the SPM (\cref{eq:D_perp_B}) with the reference MSD value $D^{\mathrm{H}_2\mathrm{O}}_{\mathrm{MSD}}$ obtained using the GROMACS tool.
    The estimate converges with only an error of $\SI{5}{\percent}$ for large $L$ but drifts off to underestimate $D$ for small $L$. (see appendix \ref{app:error_estimates} for notes on error estimates)
    }
    \label{fig:water_bulk_benchmark}
\end{figure}

To validate the SPM, we start extracting diffusion constants from trajectories sampled in molecular dynamics simulations of a homogeneous and isotropic liquid such as water. 
In such a system, standard techniques based for example on the MSD can provide a reference value $D_\mathrm{MSD}$ with excellent accuracy. 
Furthermore, an H$_2$O molecule is sufficiently small for the basic premises of the SPM to be satisfied.
Hence, these simulations are the ideal system for evaluating the SPM’s performance.

All our simulations are performed in GROMACS, by building a cubic box of a side length of $\SI{8.9}{\nano\meter}$ with a total of $23419$ SPC/E molecules (see Appendix \ref{app:simulation_methods} for full simulation details). 
After performing an equilibration protocol, a production run is performed in the NVT ensemble with periodic boundary conditions for a total of \SI{10}{\nano\second}.  
The diffusion constant $D_{\mathrm{MSD}}^{\mathrm{H}_2\mathrm{O}}$ is obtained from the mean square displacement averaged over all spatial directions and all molecules in the system throughout the entire production run using a standard GROMACS tool. 
This analysis yields a reference value of $$D_{\mathrm{MSD}}^{\mathrm{H}_2\mathrm{O}}=\SI{2.42\pm 0.01e-5}{\centi\meter\squared\per\second}.$$

To calculate the predictions of the SPM, non-overlapping and adjacent slices are chosen with a fixed slice thickness $L$, covering the entire simulation box. 
We then measure the lifetime distribution of water molecules in each slab, and calculate the average $\tau(L)$ for that slab (\cref{fig:water_bulk_benchmark}). 
The diffusion constant $D_\mathrm{SPM}$ is calculated using  \cref{eq:D_perp_B} in each slice independently. 
All obtained values are averaged to yield $D_{\mathrm{SPM}}^{\mathrm{H_2O}}(L)$ and its standard deviation for comparison with $D_\mathrm{MSD}$. 
This procedure is repeated for a range of $L$ to test the sensitivity of the SPM to the slab thickness (inset of \cref{fig:water_bulk_benchmark}). 

At high resolutions, i.e. slab thicknesses smaller than the water molecule itself ($L<\SI{0.3}{\nano\meter}$), the accuracy of  $D_\mathrm{SPM}^{\mathrm{H}_2\mathrm{O}}$ is gradually diminished. 
We attribute this to $\tau$ being comparable to the output frequency of the simulation, which statistically overestimates relatively short escape times due to discretisation errors. 
Also, on these time scales a ballistic regime appears before frequent particle-particle interactions and diffusion kicks in \cite{zwanzig1970hydrodynamic}. 
Both of these effects make the deviation from the model assumptions of purely diffusive displacement larger. 
The overall result here is an overestimation of $\tau$ and an under-estimation of $D_\perp$. 
At very low resolutions and thick slabs ($L>\SI{1.5}{\nano\meter}$), sampling the full distribution of escape times again becomes a challenge. 
The reason is that it may take a very long time for a molecule to leave the slab, the observation of which may be limited by the finite simulation time. 

At optimum resolutions, in the intermediate range $L\approx\SIrange{0.3}{1.5}{\nano\meter}$ $D_{\mathrm{SPM}}^{\mathrm{H}_2\mathrm{O}}$ is basically independent of the slab thickness, as expected. 
The obtained $$D_{\mathrm{SPM}}\approx\SI{2.5e-5}{\centi\meter\squared\per\second}$$ is only off by about $\SI{6}{\percent}$ relative to the reference value $D_{\mathrm{MSD}}^{\mathrm{H}_2\mathrm{O}}$(see \cref{fig:water_bulk_benchmark}). 
One may initially consider these systematic deviations to be a consequence of finite size effects, which have been proven to influence MSD-based diffusion results in small scale systems \cite{dunweg1993molecular}, but existing literature has actually shown that the value of the diffusion coefficient $D$ itself is affected by system size and not only one singular method of derivation \cite{yeh2004system}.  
Finite-size correction thus has to be applied equally to the SPM and the MSD results, not explaining the observed difference. 
Instead, we suspect the origin of the deviation to lie in anisotropic small scale structuring effects of the pure bulk liquid. 
Hence, the full-system MSD, which takes into account all directions as well as more data points, is less susceptible to this effect whereas we observe the SPM to be slightly more affected.

We conclude that for water, the SPM has generally proven accurate at quantitatively recovering the expected diffusion coefficient values since the water molecules are well represented by rigid, point-like particles, satisfying the underlying assumptions of the SPM method. 
Despite a small systematic over-estimation of the diffusion constant at optimal resolutions, the SPM provides absolute estimates, significantly improving on the technique employed by Bourg et. al. \cite{bourg2011molecular}.
The presented analysis, however shows that it is important to make an adequate choice of $L$. 
The resolution with which the diffusion constant can be properly determined depends on the time-step with which the trajectories are recorded (i.e. the time scale at which diffusive regime can be sampled), the order of magnitude of $D_\perp$, the internal particle dynamics, and the total simulation time.

\section{Anisotropic diffusion of liquids in confined geometries - water in a slit alumina pore}
Building on its validation on bulk liquids with a constant density background, we now employ the SPM to study systems where the diffusivity is much harder to determine. 
One such example are strongly confined liquids in nanopores.
Here, interface-adjacent dynamics as well as bulk-behaviour in direct confinement are properties that lend themselves to experimental analysis whereas simulations, in theory, allow for a more thorough investigation of the transitional region in between. 
The difficulties arise from the lack of a clear separation of length scales between the molecular size of diffusing particles, the thickness of the pore/film, and the effective interaction potentials between diffusing particles and the solid phase \cite{marion2021water}. 
These potential interactions close to interfaces may require a large number of slabs to be simulated before properties of a bulk liquid are restored \cite{prakash2017anisotropic,baer2022modelling}.

An additional issue is the geometry of the systems which induces anisotropy of the diffusion constants parallel and perpendicular to the interfaces.
This, so far, has been very challenging to characterise both experimentally and in simulations.  
Most attempts at this have employed Einstein/MSD based techniques and applied them to e.g. obtain an average second-order diffusion tensor across the entire system \cite{prakash2017anisotropic}.
Alternatively, they needed to significantly restrict the spatial resolution of their analysis to obtain reasonable locally-confined trajectories \cite{gentile2015hindered}. 
This can now be circumvented with the SPM approach which permits resolving anisotropic diffusivities close to interfaces but also in the transitional region between the pore wall and the central part of the pore. 
We will additionally employ the SPM+d approach to quantify the reliability and significance of the SPM analysis close to interfaces, where interface-particle interactions may have a major impact on the resulting diffusivity profile perpendicular to the pore wall.

\subsection{MD simulations and the SPM}

\begin{figure}
    \centering
     \includegraphics[width=0.45\textwidth]{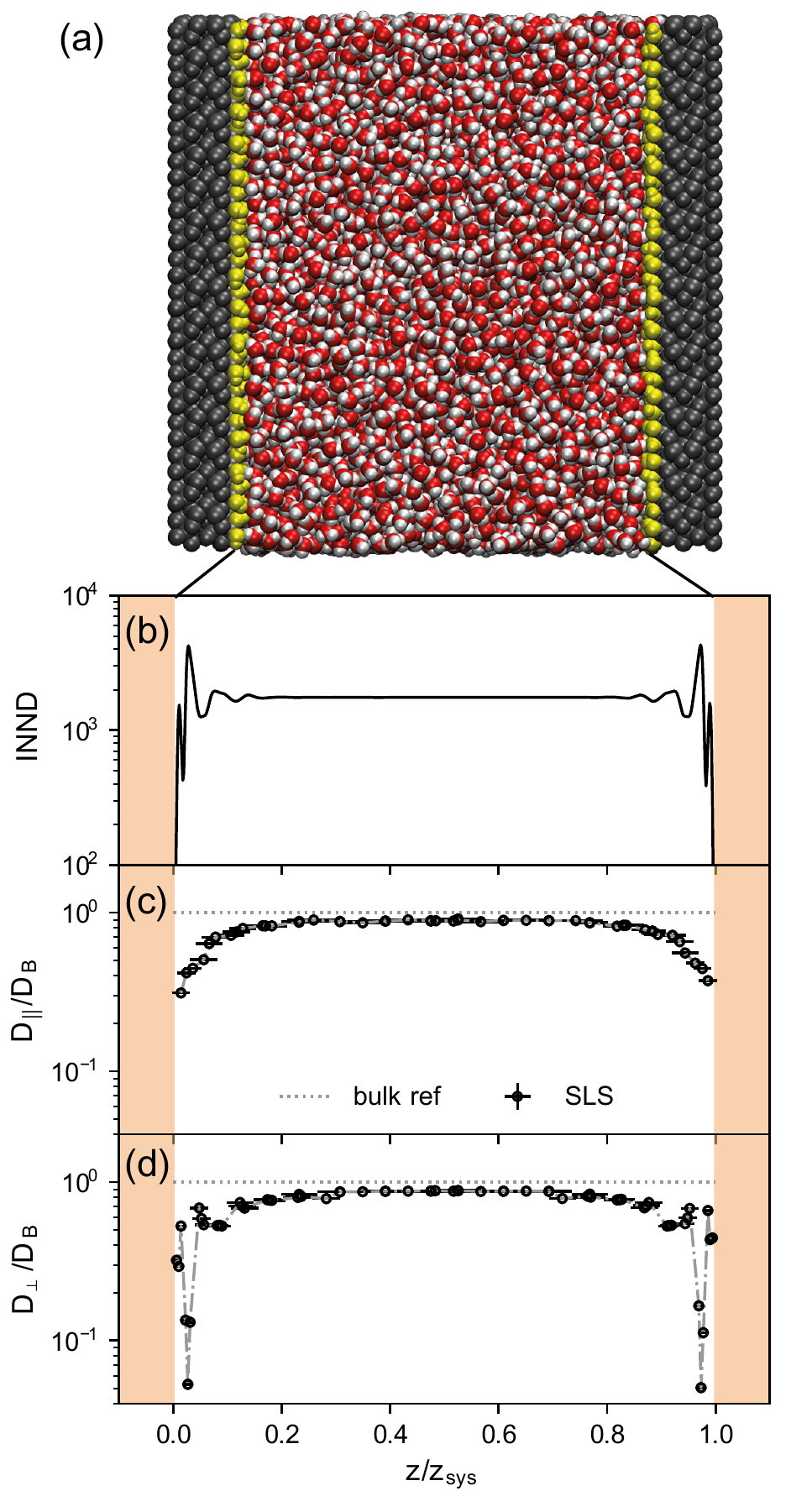} 
    \caption{\textbf{Particle transport in a water slit pore:} (a) Pore geometry employed for the water SLS simulation. The solid support (grey) with its hydroxilation (yellow) is mirrored to both sides of the liquid phase (white/red), so that confinement is created between two parallel solid layers. In all other directions, periodic boundary conditions are applied. 
    Within this slit pore, we obtain plots of (b) interface normal density profile, (c) MSD-based interface-parallel diffusion coefficient profile and (d) perpendicular lifetime based diffusion obtained via the SPM in a solid-liquid-solid pore system filled with water.
    The pore has an inner diameter $z_\mathrm{sys}\approx \SI{6}{\nano\meter}$ and the slice thicknesses are chosen in the range \SIrange{0.2}{0.5}{\nano\meter} for the MSD, and \SIrange{0.1}{0.5}{\nano\meter} for the SPM.}
    \label{fig:water_sls_resolution_benchmark}
\end{figure}

We base our investigation on MD simulations of water in a slit hydroxylated alumina pore (\cref{fig:water_sls_resolution_benchmark}a). 
For this purpose, a \SI{6}{\nano\meter} symmetric pore is filled with about $10 000$ SPC/E water molecules and equilibrated following an established protocol \cite{baer2022modelling}.  
The diffusion data are sampled over a \SI{10}{\nano\second} production run (see appendix \ref{app:simulation_methods} for methodological details). 
For the purpose of analysing diffusivity, the pore is sliced in parallel to the pore walls (see \cref{fig:scheme_system}). 
The diffusivity parallel to the pore walls ($D_\parallel$(z)) is calculated from the MSD, the latter constructed from single component trajectories over the two coordinates parallel to the wall of each molecule, sampled as long as it remains in the slab.  
The perpendicular diffusion coefficients $D_\perp(z)$ are extracted using the SPM. 

The $z$-position of diffusion coefficients is chosen to be the centre $z$-coordinate of the slice interval $[z_i,z_i+L]$ with the error bar in $x$ direction being half the slice thickness $L$, which is variable. 
Namely, the perpendicular particle mobility is expected to be lower and lifetime consequently higher at the liquid-solid-interfaces \cite{feitosa1991wall}. 
We are, therefore, able to reduce the slice thickness and increase the resolution of the SPM up to $L=\SI{0.1}{\nano\meter}$, without loss of accuracy at the solid-liquid interface. 
For the MSD, the maximal resolution is $L=\SI{0.2}{\nano\meter}$ since at smaller $L$ a linear regime is no longer observed. 
Towards the bulk, the slice thickness is increased to $L=\SI{0.5}{\nano\meter}$ due to the expected higher particle mobility comparable to the bulk system, where the larger $L$ proved necessary and reasonable. 
Also no significant variation in $D$ is expected in the centre of the pore.

The obtained $D_\parallel$ and $D_\perp$ (\cref{fig:water_sls_resolution_benchmark} c and d) are compared with the interface normal number density (INND) of atoms in the water molecules (\cref{fig:water_sls_resolution_benchmark} b). 
The diffusivity profiles in \cref{fig:water_sls_resolution_benchmark} are normalised by the reference bulk diffusivity $D_\mathrm{MSD}^{\mathrm{H}_2\mathrm{O}}$. 
To account for finite size effects, $D_\mathrm{MSD}^{\mathrm{H}_2\mathrm{O}}$ is evaluated in a bulk water system that has a similar size as the extent of SPC/E water in the pore in $x$ and $y$ direction (i.e. $(\SI{3}{\nano\meter})^3$, simulated with periodic boundary conditions). 

Focusing on the central region of the pore ($0.2 < z/z_{sys} < 0.8$), we observe a structurally bulk-like region as confirmed by comparing density correlation functions of bulk and confined water in these slabs. 
This agrees with similar findings from related studies on water in nano-channels of a similar size \cite{tsimpanogiannis2019self}.
Here, the MSD method provides values of $D_\parallel$ in good agreement with the reference bulk diffusion, mobility being only slightly lowered. 
Contrary to that, the SPM now predicts a $D_\perp$ below the bulk reference value. 
This indicates a stronger influence of the solid interfaces on perpendicular particle motion with the effect actually spanning the entirety of the pore. 
Given the structural similarity of the centre region to the bulk system, we hypothesise that this decrease originates from long-range hydrodynamic effects.

\begin{figure}
    \centering
     \includegraphics[width=0.45\textwidth]{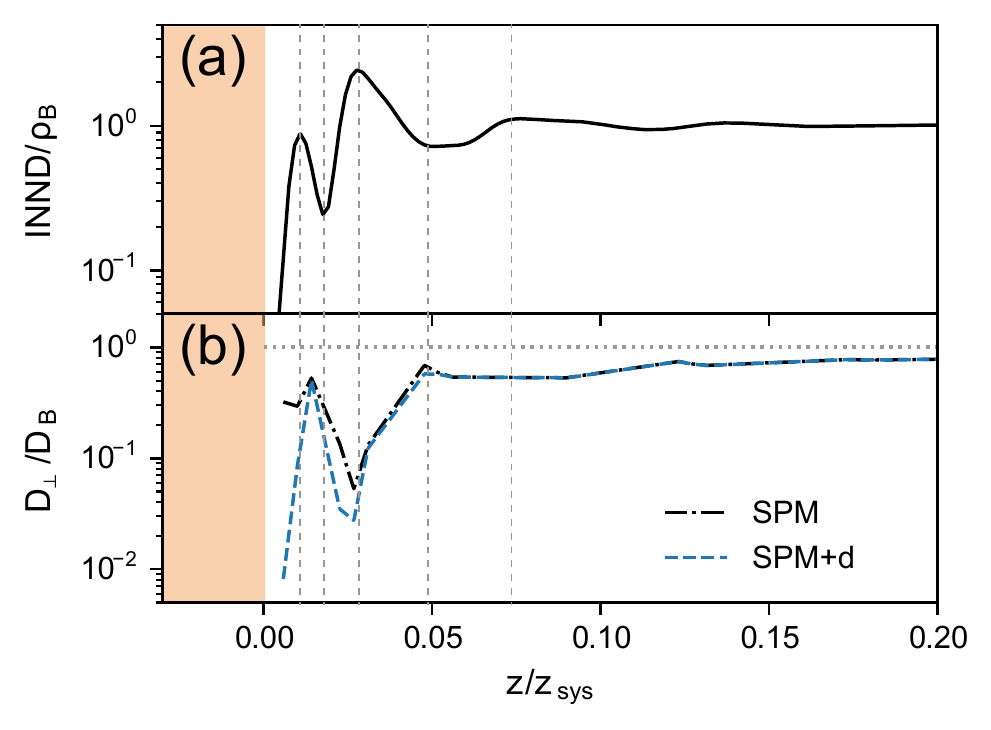} 
    \caption{\textbf{Correcting for statistical drift in a water slit pore:} (a) Interface normal density profile normalized by bulk density close to one of the SL interaces, (b) perpendicular lifetime based diffusion obtained via the SPM (dashed-dotted/black) and the first-order correction of the SPM+d (dashed/blue) in a solid-liquid-solid pore system filled with water as in \cref{fig:water_sls_resolution_benchmark}.
    Vertical lines are drawn to guide the eye at maxima and minmia of the INND.
    The impact of the drift-correction is only significant in the \SI{5}{\percent} of the film closest to the pore wall.
    } 
    \label{fig:water_sls_correction}
\end{figure}

Close to the interfaces ($0.025<z/z_{sys}<0.2$ with the lower bound signifying the position of the maximum in the INND; also the symmetrically positioned region), the difference between parallel and perpendicular diffusion is vast (\cref{fig:water_sls_resolution_benchmark} c and d).
Along the parallel direction, $D_\parallel$ drops at most by a factor of three compared to the centre region, while $D_\perp$ is up to $16$ times smaller.
The drop being more severe in $D_\perp$ due to the existence of a boundary has previously been predicted theoretically using hydrodynamic modelling approaches  \cite{bevan2000hindered,feitosa1991wall,goldman1967slow,brenner1961slow} but also measured experimentally \cite{gentile2015hindered}.
Where the parallel profile is gradual and smooth, in perpendicular direction, $D_\perp$ develops a very structured profile decaying over a region twice as thick as for $D_\parallel$. 
These fluctuations anti-correlate with the density fluctuations which can be seen by comparing INND and $D_\perp (z)$ in \cref{fig:water_sls_correction}. 
This result suggests that in this region, molecular crowding, which has been shown to suppress the in-plane (parallel) diffusivity \cite{cvitkovic2022crowding}, affects the perpendicular diffusion more than the parallel component. 
This difference in the level of structuring between $D_\perp$ and $D_\parallel$ cannot be attributed solely to the difference in resolution. 

The most intriguing result, however, is the nontrivial interplay between the water density (i.e. the effective potential) and the diffusivity perpendicular to the pore wall in the slabs close to the interfaces ($0<z/z_{sys}<0.025$ as well as on the other interface). 
In this region (see \cref{fig:water_sls_correction}), water can form hydrogen bonds with the hydroxyl groups on the surface of the alumina pore. 
This similarly affects both components of diffusivity. 
Importantly, however, in this region, the interaction potential with the surface is also oscillating, with steep gradients. 
Therefore, it must be confirmed that these observations are not a consequence of the assumption of a constant potential within each slice (albeit of different amplitude), as imposed by the SPM.

\subsection{Estimate of the systematic error by explicitly accounting for drift}

Equipped with the previously presented SPM+d approach, which accounts for the impact of a non-constant drift term on the results of the diffusivity analysis, we are able to validate the significance of these observed oscillations in $D_\perp$. 
To extract the necessary ratio $\mu/D$ for the calculation of the correction coefficient $K(\gamma)$ in the SPM+d approach, we fit the logarithmic density profile within a slice with a linear function and use the resulting slope as the $\mu/D$ value for $\gamma$ (see appendix \ref{app:theory_spmd}). 
We then use the known slice thickness $L$ to calculate $\gamma=\mu L/D$ and the resulting correction $K(\gamma)$ to arrive at the diffusion coefficient $D_{\perp,+d}$.

Plotting this new (SPM+d) profile in tandem with the profile according to the SPM (see \cref{fig:water_sls_correction}b), we actually observe virtually no change of the diffusion profile except for the \SI{5}{\percent} of the film immediately on top of the solids. 
There, the correction as a consequence of the first-order perturbation theory presented in the SPM+d actually leads to a further reduction of the diffusivitiy values in the dips and thus a more pronounced oscillation profile than through the SPM alone.  
Consequently, we arrive at the conclusion that the characteristic profile of interface-perpendicular mobility changes in the water filled pore are actual changes in particle mobility as a consequence of the density oscillations associated with the formation of solvation layers. 
While these results provide fundamentally new insights into the anisotropic mobility of water in narrow pores, they also demonstrate the power of the SPM and SPM+d approaches when applied carefully in an appropriate system.

\section{Discussion and Conclusions}
In this work, we have introduced a novel, so-called SPM technique for quantitatively analysing anisotropic diffusion of small diffusing objects with an extension to explicitly account for the effect of local drift in confined geometries.
Our SPM results are obtained without the need for prior calibration as required for alternative, so-far available approaches \cite{bourg2011molecular}. 
The prowess of this approach has been demonstrated on the paradigmatic select case of bulk water as well as a water phase in a slit pore. 
We demonstrate that the SPM can quantitatively reproduce key aspects of diffusion by finding agreement with theoretical predictions \cite{goldman1967slow,brenner1961slow}, simulations \cite{prakash2017anisotropic} and experimental measurements \cite{gentile2015hindered}. 

The SPM has --- by design --- opened up new opportunities to resolve a standing problem \cite{tsimpanogiannis2019self} of characterising the anisotropic diffusion tensor of liquids in confinement, where the conditions for the application of methods, such as the Einstein/MSD \cite{uhlenbeck1930theory} and the Green--Kubo  \cite{kubo1957statistical,green1954markoff} approaches, are not met. 
However, due to the dependence of the SPM approach on accurate lifetime statistics, significant care needs to be put towards influences that could alter these results.
Besides taking care of performing the calculations at a reasonable resolution, the most obvious influence on SPM results could emerge from an effective interface-normal background potential. 
The latter is expected to introduce drift within the subspace of interest, causing lifetime statistics to be under- and diffusivity to be overestimated by the SPM. 
We explicitly capture this effect in the SPM+d approach that delivers an analytic solution to the simplified Smoluchowski equation with constant non-vanishing drift.
Notably, the error introduced by the drift can be controlled by the choice of resolution in the SPM approach, rendering it useful even close to interfaces, where the gradients of the underlying potential may be large. 
This allows for a more thorough analysis of interface-adjacent diffusion profiles that has so far not been possible. 

Further improvement of the SPM+d approach could be achieved through the calculation of the correction factor of the diffusivity from to the skewness of escape probabilities to either boundary of the subspace, as discussed recently in related literature \cite{belousov2022statistical}. 
In doing so, the extended approach would at the same time allow for the analysis of the local free energy surface experienced by a diffusing object \cite{hummer2005position,rosta2015free}. 
As such, information on the diffusivity becomes accessible even if the potentials cannot be fully resolved. 
For example, in experiments, the effective potential between the particle and the wall may not be readily available, unlike in MD simulations.

The potential of the SPM and SPM+d approaches is best demonstrated by our test case of water confined to a slit pore. 
Based on the presented first-order correction theory, we show the strong anisotropy of water diffusion parallel and perpendicular to the pore wall in the interfacial layers. 
Furthermore, we are able to demonstrate the significance of observed oscillations in perpendicular diffusivity which is related to the interface normal density profile in a more complex manner than expected. 

The main issue with the SPM and SPM+d approaches is the requirement that the diffusive particle is small and possesses no relevant internal dynamics.
This, of course, limits their application to only simple particles and liquids. 
More complex molecular liquids and larger flexible particles do not satisfy these conditions. 
Nevertheless, the SPM approach can be enriched to account for the internal degrees of freedom of diffusing species coupling to translations, as presented in the follow-up work \cite{epm_pub}.

\printcredits

\section{Acknowledgements}
The authors declare no competing financial interest.

We acknowledge funding by the Deutsche Forschungsgemeinschaft (DFG, German Research Foundation) – Project-ID 416229255 – SFB 1411 Particle Design and - Project-ID 431791331 - SFB 1452 Catalysis at Liquid Interfaces (CLINT).
The authors gratefully acknowledge the scientific support and HPC resources provided by the Erlangen National High Performance Computing Center (NHR@FAU) of the Friedrich-Alexander-Universit\"{a}t Erlangen-N\"{u}rnberg (FAU).

\appendix

\renewcommand\thefigure{A\arabic{figure}}%
\renewcommand\thetable{A\arabic{table}}%

%
%
%
%
%
%

\section{Simple particle Model} \label{app:theory_spm}
Assuming a point-like particle diffusing purely with no external force field, we can model the diffusive process using the Smoluchowski equation:
\begin{align}
   \partial_t \rho(\vec{x},t) &= \partial_x \left(D(\vec{x})\partial_x\rho(\vec{x},t)\right)
\end{align}
where $D$ denotes the local diffusion coefficient and $\rho$ the local particle density.
As we want to derive a link between the mean time of a particle's stay in a confined subspace, we will think of $\rho$ in terms of a probability density representing the probability that the particle is at time $t$ at position $\vec{x}$ without having left the subspace in between time $0$ and time $t$.
Integrating over the entire subspace at time $t$ will give us the overall probability $p(t)$ of a particle remaining within the confines of the subspace, which will go down from $p(0)=1$ to $p(t\to\infty)\to 0$ due to the particles diffusing out of the subspace. 
To model an overall mean evolution, we will assume an initial uniform distribution of the probability density over the subspace at $t=0$. 
Additionally, we assume isotropy and translational symmetry of the subspace along all but one axes, so we can integrate over those and only consider the density distribution along one axis. This also enables us to represent the slice of space considered as our confined subspace as an interval $[0,L]$ of thickness $L$ along the last remaining axis, which we denote as the $z$-axis, with absorbing boundary conditions, modelling the particle leaving the subspace. 

Hence, we simplify above equation to the following system of PDE and initial/boundary conditions:
\begin{align}
    \partial_t \rho(z,t) &=\partial_z\left(D(z)\partial_x \rho(z,t)\right),&t\in[0,\infty), z\in[0,L]\\
    \rho(z,0) &= p_0,\hspace{20pt} &z\in(0,L)\\
    \rho(0,t) &= \rho(L,t) = 0,&t\in[0,\infty).
\end{align}
To solve this PDE, we will further assume that $D(z)$ be sufficiently constant across the extent of the interval $I_L = [0,L]$ due to the background density of all particles (not just those not having left the slab between $0$ and $t$) remaining statistically constant, thus enabling us to solve the equation by finding the eigenfunctions of the equation:
\begin{align}
    -\lambda\rho &= D\partial_z^2 \rho
\end{align}
which in general amount to linear combinations of sine and cosine curves, but due to the absorbing boundary conditions are limited to sine-curves of the shape:
\begin{align}
    f_n(z) = c_n \sin\left(\frac{n\pi}{L} z\right)
\end{align}
where 
\begin{align}
    c_n = \left(\int_0^L\sin^2\left(\frac{n\pi}{L} z\right)\,dz\right)^{-\frac{1}{2}}
\end{align}
is a normalisation coefficient and $\lambda_n = D\left(\frac{n\pi}{L}\right)^2$ is the eigenvalue.
We then need to decompose the initial condition in terms of a scalar product for which we will choose the default scalar product for real-valued functions on the interval $I_L$:
\begin{align}\langle f,g\rangle = \int_0^L f(z) g(z) \,dz\end{align}
This provides us with decomposition coefficients:
\begin{align}a_n = \langle f_n,p_0\rangle\end{align}
and fixes $c_n$ such that:
\begin{align}\langle f_n, f_n\rangle =1 \Leftrightarrow c_n^2 = \frac{1}{\int_0^L\sin^2\left(\frac{n\pi}{L} z\right)\,dz}\end{align}
so that overall:
\begin{align}\rho(z,t) = \sum_{n=1}^\infty a_n  f_n(z)  \exp\left(-\lambda_n t\right)\end{align}
Integrating over $I_L$ (or the scalar product with the constant function $I(z)=1$) then yields the survival probability $p(t)$:
\begin{align}
    p(t) &= \langle \rho(\cdot,t),I\rangle = \sum_{n=1}^\infty a_n  \langle f_n(z),I\rangle  \exp\left(-\lambda_n t\right)
\end{align}
From probability theory for purely non-negative random variables, we know that integrating $p(t)$ from $t=0$ to $t=\infty$ will yield the average survival/crossing lifetime $\tau$ of the particle:
\begin{align}
   \tau &= \int_0^\infty p(t)\,dt \\
   &= \sum_{n=1}^\infty a_n  \langle f_n(z),I\rangle \int_0^\infty  \exp\left(-\lambda_n t\right)\,dt\\
   &= \sum_{n=1}^\infty a_n  \langle f_n(z),I\rangle \frac{1}{\lambda_n}\\
   &= \sum_{n=1}^\infty \langle f_n(z),p_0 I\rangle  \langle f_n(z),I\rangle \frac{1}{\lambda_n}\\
   &= \sum_{n=1}^\infty p_0 \langle f_n(z),I\rangle  \langle f_n(z),I\rangle \frac{1}{\lambda_n}\\
   &= \sum_{n=1}^\infty p_0  \left(\langle f_n(z),I\rangle\right)^2 \frac{1}{\lambda_n}\\
   &= \sum_{n=1}^\infty \underbrace {p_0}_{=\frac{1}{L}}  \frac{1}{\lambda_n} c_n^2\left(\int_0^L\sin\left(\frac{n\pi}{L} z\right)\,dz \right)^2\\
   &= \frac{1}{L} \sum_{n=1}^\infty \frac{1}{\lambda_n} \frac{\left(\int_0^L\sin\left(\frac{n\pi}{L} z\right)\,dz \right)^2}{\int_0^L\sin^2\left(\frac{n\pi}{L} z\right)\,dz}\\
   &= \frac{1}{L} \sum_{n=1}^\infty \frac{1}{\lambda_n} \frac{\left(\frac{L}{\pi n} \left(1-\cos(\pi n)\right)\right)^2}{\frac{L}{2}}\\
   &= \frac{1}{L} \sum_{n=1}^\infty \frac{1}{D}\left(\frac{L}{n\pi}\right)^2 \frac{\left(\frac{L}{\pi n} \left(1-(-1)^n\right)\right)^2}{\frac{L}{2}}\\
   &= \frac{2L^2}{D\pi^4} \sum_{n=1}^\infty \frac{1}{n^4} \left(1-(-1)^n\right)^2\\
   &= \frac{8L^2}{D\pi^4} \underbrace{\sum_{n=0}^\infty \frac{1}{(2n+1)^4}}_{\text{Can be calculated from $\zeta(4)$}}\label{eq:use_of_zeta}\\
   &= \frac{8L^2}{D\pi^4}\times \frac{\pi^4}{96} \\
   &= \frac{1}{12}\frac{L^2}{D}
\end{align}
Or, rearranging above equation: 
\begin{align}
    D_\mathrm{B} &= \frac{1}{12} \frac{L^2}{\tau} \label{eq:default_bulk_connection}
\end{align}
which leaves us with the result presented in the main manuscript (\cref{eq:D_perp_B}).

Before, in \cref{eq:use_of_zeta}, we used the fact, that:
\begin{align}
    \sum_{n=0}^\infty \frac{1}{(2n+1)^4} &= \sum_{n=1}^\infty \frac{1}{n^4} - \sum_{n=1}^\infty \frac{1}{(2n)^4}\\
    &=\sum_{n=1}^\infty \frac{1}{n^4} - 2^{-4}\sum_{n=1}^\infty \frac{1}{n^4}\\
    &= \frac{15}{16}\sum_{n=1}^\infty \frac{1}{n^4}  =\frac{15}{16}  \zeta(4)= \frac{15}{16}\times \frac{\pi^4}{90} = \frac{\pi^4}{96} \label{eq:zeta_identity}
\end{align}

One can easily see, that the assumptions of the above model do not hold for the scenario of an interface slab, where at least one of the interval boundaries is not absorbing but instead a reflective boundary through which the particle cannot leave the subspace. 
Let this without loss of generality be at $z=0$.
In terms of the model, this replaces the condition 
\begin{align}
\rho(0,t) = 0,\hspace{20pt} t\in[0,\infty)
\end{align}
with 
\begin{align}
\partial_z\rho(0,t) = 0,\hspace{20pt} t\in[0,\infty)
\end{align}
leaving us with eigenfunctions 
\begin{align}
f_n(z) = c_n \cos\left(\frac{(2n+1)\pi}{2L} z\right)
\end{align}
where 
\begin{align}
    c_n = \left(\int_0^L\cos^2\left(\frac{(2n+1)\pi}{2L} z\right)\,dz\right)^{-\frac{1}{2}}
\end{align}
is again a normalisation coefficient and their eigenvalues become $\lambda_n = D\left(\frac{(2n+1)\pi}{2L}\right)^2$.
Performing the same procedure as above, we again arrive at
\begin{align}
    p(t) &= \langle \rho(\cdot,t),I\rangle = \sum_{n=1}^\infty a_n  \langle f_n(z),I\rangle  \exp\left(-\lambda_n t\right)
\end{align}
thus resulting in the new estimate for the mean lifetime:
\begin{align}
   \tau &= \int_0^\infty p(t)\,dt \\
   &= \sum_{n=1}^\infty a_n  \langle f_n(z),I\rangle \int_0^\infty  \exp\left(-\lambda_n t\right)\,dt\\
   &= \sum_{n=1}^\infty a_n  \langle f_n(z),I\rangle \frac{1}{\lambda_n}\\
   &= \sum_{n=1}^\infty \langle f_n(z),p_0 I\rangle  \langle f_n(z),I\rangle \frac{1}{\lambda_n}\\
   &= \sum_{n=1}^\infty p_0 \langle f_n(z),I\rangle  \langle f_n(z),I\rangle \frac{1}{\lambda_n}\\
   &= \sum_{n=1}^\infty p_0  \left(\langle f_n(z),I\rangle\right)^2 \frac{1}{\lambda_n}\\
   &= \sum_{n=1}^\infty \underbrace {p_0}_{=\frac{1}{L}}  \frac{1}{\lambda_n} c_n^2\left(\int_0^L\cos\left(\frac{(2n+1)\pi}{2L} z\right)\,dz \right)^2\\
   &= \frac{1}{L} \sum_{n=1}^\infty \frac{1}{\lambda_n} \frac{\left(\int_0^L\cos\left(\frac{(2n+1)\pi}{2L} z\right)\,dz \right)^2}{\int_0^L\cos^2\left(\frac{(2n+1)\pi}{2L} z\right)\,dz}\\
   &= \frac{1}{L} \sum_{n=1}^\infty \frac{1}{\lambda_n} \frac{\left(\frac{2L}{\pi (2n+1)}\cos(\pi n)\right)^2}{\frac{L}{2}}\\
   &= \frac{1}{L} \sum_{n=1}^\infty \frac{1}{D}\left(\frac{2L}{(2n+1)\pi}\right)^2 \frac{\left(\frac{2L(-1)^n}{\pi (2n+1)}\right)^2}{\frac{L}{2}}\\
   &= \frac{32L^2}{D\pi^4}  \underbrace{\sum_{n=0}^\infty \frac{1}{(2n+1)^4}}_{\text{again using \cref{eq:zeta_identity}}}\\
   &= \frac{32L^2}{D\pi^4}\cdot \frac{\pi^4}{96}  =\frac{1}{3}\frac{L^2}{D}
\end{align}
This leaves us with four times the expected mean lifetime of a bulk-like slab with the same diffusion constant and same thickness $L$ for interface slabs or -- rearranging:
\begin{align}
    D_\mathrm{LV} &= \frac{1}{3} \frac{L^2}{\tau} \label{eq:default_interface_connection}
\end{align}
as an estimator for the mean diffusion coefficient.

Notably, alternative, and perhaps simpler methods could be used to derive \cref{eq:default_interface_connection,eq:default_bulk_connection} \cite{berg1993random,bicout1998turnover,van2007water,mercier2016diffusion}, the here used explicit decomposition into eigenfunctions is chosen because it can be  easily adapted to the more  complex  situations  of a non-constant effective potential/free energy surfaces across the slab, which is a natural extension of the presented model explicitly presented in appendix \ref{app:theory_spmd}.

%
%
%
%
%
%

\section{Effect of drift on the simple particle model}
\label{app:theory_spmd}
In terms of the Smoluchowski equation used for the derivation of the SPM, we will try and solve it for the simplest scenario involving the presence of a none-zero drift term, i.e. constant drift $\mu=const.$, for a bulk-like slab to estimate the error introduced by our prior omission of the drift term.

We therefore attempt to construct the solutions to the following simplified equation:
\begin{align}
    \partial_z p =&  -\mu p' + D p''
\end{align}
We can construct the set of eigenfunctions:
\begin{align}
    -\lambda_n p_n=& -\mu p_n'+Dp_n''\\
    p_n=& e^{\frac{\mu}{2D}z}\sin\left(\frac{n\pi}{L}z\right)
\end{align}
whose eigenvalues are --- in agreement with the SPM for $\mu=0$ --- shifted to 
\begin{align}
    \lambda_n =  \frac{\mu^2}{4D} +D\left(\frac{n\pi}{L}\right)^2
\end{align}
and the eigenfunctions turn into:
\begin{align}
    p_n=& e^{\frac{\mu}{2D} z}\sin\left(\frac{n\pi}{L}z\right).
\end{align}
We calculate:
\begin{align}
    \langle f_n,f_n\rangle =& \int_0^L  e^{\frac{\mu}{D} z}\sin^2\left(\frac{n\pi}{L}z\right) \,dz\\
    =& \frac{D}{2\mu}\frac{n^2\left(e^{\frac{\mu L}{D}}-1\right)}{n^2+\frac{\mu^2 L^2}{4\pi^2 D^2}}
\end{align}
as well as
\begin{align}
    \langle 1,f_n\rangle =& \int_0^L  e^{\frac{\mu}{2D} z}\sin\left(\frac{n\pi}{L}z\right) \,dz\\
    =&\frac{4\pi D^2 L n\left(1-(-1)^n \exp\left(\frac{L\mu}{2D}\right) \right)}{4\pi^2D^2n^2+L^2\mu^2}
\end{align}
arriving at:
\begin{align}
\tau &= \int_0^\infty p(t)\,dt \\
   &= \sum_{n=1}^\infty \frac{\langle f_n(z),p_0(z)\rangle  \langle f_n(z),I\rangle}{\langle f_n(z),f_n(z)\rangle} \frac{1}{\lambda_n} \label{eq:raw_drift_time_series}
\end{align}
For the potential and initial conditions, we have:
\begin{align}
    U = -kT \ln(\rho) \Longleftrightarrow &\rho = \exp(- \beta U)\\
    \mu =& \nu \cdot F = -\nu \partial_z U\\
    U =& -\frac{\mu}{\nu}z
\end{align}
where $\mu$ is the constant drift, $\nu$ is the particle mobility, $k_B$ is the Boltzman constant and $T$ is the absolute temperature with $\beta = 1/k_BT$.
We also know from the Einstein relation that $D= k_BT\nu$, hence:
\begin{align}
    p_0=& c\exp\left(-\frac{\mu z}{D}\right)\\
    c =&\left[\frac{D}{\mu}\left(1-\exp\left(-\frac{\mu L}{D}\right)\right)\right]^{-1}
\end{align}
Consequently:
\begin{align}
    \langle p_0,f_n\rangle =& c\int_0^L  e^{-\frac{\mu}{2D} z}\sin\left(\frac{n\pi}{L}z\right) \,dz\\
    =&c\frac{4\pi D^2 L n\left(1-(-1)^n \exp\left(-\frac{L\mu}{2D}\right) \right)}{4\pi^2D^2n^2+L^2\mu^2}
\end{align}
Plugging all individual results into \cref{eq:raw_drift_time_series}, after some calculation we arrive at:
\begin{align}
\tau 
   &= \frac{2}{\pi^4}\frac{L^2}{D}\frac{\frac{\mu^2 L^2}{D^2}}{ \cosh\left(\frac{\mu L}{D}\right)-1 } \sum_{n=1}^\infty \frac{ \left(1-(-1)^n \cosh\left(\frac{L\mu}{2D}\right) \right)}{\left(n^2+\frac{\mu^2L^2}{4\pi^2 D^2} \right)^2} 
\end{align}
for which no closed analytical presentation is known to us.

We check that the convergence in the no-drift case $\mu=0$ is consistent with the SPM:
\begin{align}
    \lim_{\mu\to 0} \tau
   &= \frac{4}{\pi^4}\frac{L^2}{D} \sum_{n=1}^\infty \frac{1-(-1)^n}{n^4}\\
   &= \frac{8}{\pi^4}\frac{L^2}{D} \sum_{n=1}^\infty \frac{1}{(2n+1)^4} = \frac{8}{\pi^4}\frac{L^2}{D}\frac{\pi^4}{96}= \frac{1}{12}\frac{L^2}{D}
\end{align}
i.e., indeed, the resulting series for the drift-including model is consistent with the simple particle model.

Based on our insights into how the relative error of the observed lifetime relation evolves (\cref{fig:drift_theory_coefficient}), we can now make a better estimate of the effect of neglecting the drift.

More notably, we can denote the drift-induced relative change $K(\gamma)$:
\begin{align}
    K\left(\gamma\right) =\frac{24}{\pi^4}\frac{\gamma^2}{ \cosh\left(\gamma\right)-1 } \sum_{n=1}^\infty \frac{ \left(1-(-1)^n \cosh\left(\frac{\gamma}{2}\right) \right)}{\left(n^2+\frac{\gamma^2}{4\pi^2} \right)^2} \label{eq:app_drift_induced_correction}
\end{align}
where $\gamma=\mu L /D$ is the scale of the induced drift in relation to the diffusion coefficient $D$ and the slice thickness $L$.
We observe that $K$ is actually a function in $\gamma^2$, making the correction independent of the direction or the sign of the induced drift due to the local symmetry. 

Then, using \cref{eq:drift_induced_correction}, we can rewrite \cref{eq:default_bulk_connection} to provide us with a new relation between $L$, $D$ and $\tau$, which also accounts for the second-order correction due to $\mu$:
\begin{align}
    D_\mathrm{B}(L,\tau,\mu) =& \frac{1}{12} K\left(\frac{\mu L}{D}\right) \frac{L^2}{\tau}. \label{eq:app_drift_accounting_bulk_connection}
\end{align}
In principle, this equation appears to make the derivation of $D_\mathrm{B}$ more problematic due to the appearance of $D$ on both sides of the equation.
In fact, we can still employ this new relation to provide an estimate for how much particle mobility is affected by the density fluctuations close to interfaces. 
More specifically, we know from $\rho \propto \exp\left(\frac{\mu z}{D}\right)$, that we can obtain the ratio $\mu/D$ for the use in the argument of $K(\gamma)$ on the r.h.s by fitting the logarithmic density profile in a slab with a linear function and employing the slope of that fit function together with the chosen slice thickness $L$ to calculate $\gamma$ and the resulting $K(\gamma)$. 

We would like to point out, that according to our derivation, the first order correction of the diffusion-lifetime relation $K(\gamma)$ applied in \cref{eq:app_drift_accounting_bulk_connection} can only ever lead to a lower diffusion coefficient $D_\mathrm{B}$ than according to \cref{eq:default_bulk_connection}.
Furthermore, the error due to the omission of $\mu$ in our derivation of the SPM is dependent on the square of $L$, allowing for the analysis to limit the impact of the neglected drift term in general scenarios.

%
%
%
%
%
%

\section{Simulation methods for water systems} \label{app:simulation_methods}

The water is in all systems parameterised by the SPC/E model.
After creation of the system, an energy minimisation is performed before velocities are initialised according to a Maxwell distribution of the desired temperature of \SI{293.15}{\K}.
Subsequently, a \SI{5}{\ns} NPT equilibration run at ambient pressure is performed to adjust the system density (for details for each system see below).
A final NVT equilibration run is conducted for \SI{1}{\ns} to account for equilibration under production conditions.

In all simulations with water systems the BDP velocity rescaling thermostat \cite{Bussi2007} is used with a coupling time of \SI{1.0}{\ps}, to keep the system at \SI{293.15}{\K}.
In the NPT runs, the C-rescale barostat is used to control the pressure with a coupling time of \SI{5.0}{\ps}.
Further parameters of the simulation like cut-off radii and treatment of electrostatic interactions are the same as for the IL systems.
All water simulations are run with \texttt{GROMACS 2021.3}.

\subsection{Pure water system}
For benchmarking in the pure water system, we used two simulations of pure water in a cubic box, with $1000$ and $23419$ SPC/E water molecules, resulting in box side lengths of \SI{3.1}{\nm} and \SI{8.9}{\nm}, respectively.
The system is created from a small box of pre-equilibrated liquid water to minimise the required equilibration time.
During the NPT run, isotropic box scaling was applied to adjust the pressure in the system.
The production runs over the course of a $\SI{10}{\nano\second}$.

\subsection{Water in the slit alumina pore}
The SLS simulations with a water filled pore (see \cref{fig:water_sls_resolution_benchmark}a) use the same solid support as the IL SLV system.
However, the solid is mirrored to negative $z$-direction, to create a slit pore wrapping around periodic boundary conditions in $z$-direction.
At first, the pore is created in an empty box that accommodates a pore void slightly larger than is needed for the desired amount of water in the pore.
This void is then filled with $10431$ SPC/E water molecules and an energy minimisation is performed.
A single step of equilibration is performed in the NPT ensemble, where the box scaling only adjusted the width of the pore to obtain the correct liquid density.
The resulting system has a size of \SI{10.41}{\nano\meter} in $z$-direction orthogonal to the sapphire slabs and a resulting pore thickness of about  \SI{6}{\nano\meter} filled with water.
The production run of the water slit pore covered an additional total statistically usable time of $\SI{10}{\nano\second}$.

%
%
%
%
%
%
\section{Error estimates}\label{app:error_estimates}
In our paper, we use several differently distributed random variable statistics. 
Most notable among those are the Mean Square Displacement (MSD) of the Einstein approach.
We generally assume the MSD to be the estimated variance of a normal distributed random variable. 
Hence, the MSD is a prime example of a $\chi^2$-distributed variable for which we can provide an error estimator via the usual estimator for the variance. 

Let 
\begin{align}
    S= \frac{1}{n-1}\sum_n (X_n-\overline{X})^2
\end{align}
denote the standard unbiased estimator for the variance of a sample set of size $n$. 
Then 
$$Q=\frac{(n-1)S}{\sigma^2}$$
is expected to be $\chi^2_{n-1}$-distributed ($\chi^2$ with $n-1$ degrees of freedom) and we can use this to derive a $(1-\alpha)100\si{\percent}$-confidence interval for $\sigma^2$:

\begin{align}
    P\left(\sigma^2 \in \left[\frac{(n-1)S^2}{\chi^2_{\frac{\alpha}{2},n-1}},\frac{(n-1)S^2}{\chi^2_{1-\frac{\alpha}{2},n-1}}\right]\right) = 1-\alpha
\end{align}

where $\chi^2_{p,k}$ is defined by:
\begin{align}
    P(X > \chi^2_{p,k}) = 1-p
\end{align}
with $X$ being a $\chi^2_{k}$-distributed random variable.

Wherever an error or confidence interval for the MSD is denoted in our graphs or calculations (e.g. in linear fits for the derivation of $D$), we use this estimate of the confidence interval with $\alpha=0.05$.

For the lifetime distributions, our calculations show, that the lifetimes are distributed like an overlay of multiple exponential functions. 
To obtain a confidence interval for the mean lifetime, we thus use the mid- to long-term approximation of the lifetime being approximately exponentially distributed to derive a confidence interval.

Let $\overline{X}$ denote the mean lifetime of a sample set obtained from $n$ data points, then the $(1-\alpha)100\si{\percent}$-confidence interval for the mean lifetime $\tau$ is given by:
\begin{align}
    P\left(\tau \in \left[\frac{2n\overline{X}}{\chi^2_{\frac{\alpha}{2},2n}},\frac{2n\overline{X}}{\chi^2_{1-\frac{\alpha}{2},2n}}\right]\right) = 1-\alpha
\end{align}
Hence, this sets a confidence interval to estimate the true range of $\tau$ with $\alpha=0.05$ as well as the estimate of the resulting error of the mean diffusion coefficient $D_\perp$.

%
%
%
%
%
%

\section{Additional software resources}\label{app:provided_software}
We supply material on  the accompanying github project page \url{https://github.com/puls-group/diffusion_in_slit_pores}. 
It contains the script for the calculation of the SPM+d correction coefficient $K(\gamma)$ and tools to calculate the diffusivities based on Gromacs trajectories,
In addition to the github project, where active development may be going on and where we also invite feedback and bug reports, we also offer an archived version of the code on Zenodo \cite{hollring_kevin_2022_7446071}.

\FloatBarrier

\bibliographystyle{model1a-num-names}


\bibliography{diffusion_jcis_spm}

\providecommand{\noopsort}[1]{}\providecommand{\singleletter}[1]{#1}%
\begin{thebibliography}{78}
\expandafter\ifx\csname natexlab\endcsname\relax\def\natexlab#1{#1}\fi
\providecommand{\url}[1]{\texttt{#1}}
\providecommand{\href}[2]{#2}
\providecommand{\path}[1]{#1}
\providecommand{\DOIprefix}{doi:}
\providecommand{\ArXivprefix}{arXiv:}
\providecommand{\URLprefix}{URL: }
\providecommand{\Pubmedprefix}{pmid:}
\providecommand{\doi}[1]{\href{http://dx.doi.org/#1}{\path{#1}}}
\providecommand{\Pubmed}[1]{\href{pmid:#1}{\path{#1}}}
\providecommand{\bibinfo}[2]{#2}
\ifx\xfnm\relax \def\xfnm[#1]{\unskip,\space#1}\fi
\bibitem[{Kramers(1940)}]{kramers1940brownian}
\bibinfo{author}{H.~A. Kramers}, \bibinfo{journal}{Physica} \bibinfo{volume}{7}
  (\bibinfo{year}{1940}) \bibinfo{pages}{284--304}.
\bibitem[{Marion et~al.(2021)Marion, Vu{\v{c}}emilovi{\'c}-Alagi{\'c},
  {\v{S}}padina, Radenovi{\'c}, and Smith}]{marion2021water}
\bibinfo{author}{S.~Marion},
  \bibinfo{author}{N.~Vu{\v{c}}emilovi{\'c}-Alagi{\'c}},
  \bibinfo{author}{M.~{\v{S}}padina}, \bibinfo{author}{A.~Radenovi{\'c}},
  \bibinfo{author}{A.-S. Smith}, \bibinfo{journal}{Small} \bibinfo{volume}{17}
  (\bibinfo{year}{2021}) \bibinfo{pages}{2100777}.
\bibitem[{Lee et~al.(2017)Lee, Fenter, Nagy, and Sturchio}]{lee2017real}
\bibinfo{author}{S.~S. Lee}, \bibinfo{author}{P.~Fenter},
  \bibinfo{author}{K.~L. Nagy}, \bibinfo{author}{N.~C. Sturchio},
  \bibinfo{journal}{Nature Communications} \bibinfo{volume}{8}
  (\bibinfo{year}{2017}) \bibinfo{pages}{1--9}.
\bibitem[{Siretanu et~al.(2014)Siretanu, Ebeling, Andersson, Stipp, Philipse,
  Stuart, Van Den~Ende, and Mugele}]{siretanu2014direct}
\bibinfo{author}{I.~Siretanu}, \bibinfo{author}{D.~Ebeling},
  \bibinfo{author}{M.~P. Andersson}, \bibinfo{author}{S.~Stipp},
  \bibinfo{author}{A.~Philipse}, \bibinfo{author}{M.~C. Stuart},
  \bibinfo{author}{D.~Van Den~Ende}, \bibinfo{author}{F.~Mugele},
  \bibinfo{journal}{Scientific Reports} \bibinfo{volume}{4}
  (\bibinfo{year}{2014}) \bibinfo{pages}{1--7}.
\bibitem[{Kestin and Wakeham(1988)}]{kestin1988transport}
\bibinfo{author}{J.~Kestin}, \bibinfo{author}{W.~A. Wakeham},
  \bibinfo{title}{Transport properties of fluids: thermal conductivity,
  viscosity, and diffusion coefficient}, volume~\bibinfo{volume}{1},
  \bibinfo{publisher}{Hemisphere Publishing Corporation}, \bibinfo{year}{1988}.
\bibitem[{Lhermerout et~al.(2018)Lhermerout, Diederichs, and
  Perkin}]{lhermerout2018ionic}
\bibinfo{author}{R.~Lhermerout}, \bibinfo{author}{C.~Diederichs},
  \bibinfo{author}{S.~Perkin}, \bibinfo{journal}{Lubricants}
  \bibinfo{volume}{6} (\bibinfo{year}{2018}) \bibinfo{pages}{9}.
\bibitem[{Gebbie et~al.(2017)Gebbie, Smith, Dobbs, Warr, Banquy, Valtiner,
  Rutland, Israelachvili, Perkin, Atkin et~al.}]{gebbie2017long}
\bibinfo{author}{M.~A. Gebbie}, \bibinfo{author}{A.~M. Smith},
  \bibinfo{author}{H.~A. Dobbs}, \bibinfo{author}{G.~G. Warr},
  \bibinfo{author}{X.~Banquy}, \bibinfo{author}{M.~Valtiner},
  \bibinfo{author}{M.~W. Rutland}, \bibinfo{author}{J.~N. Israelachvili},
  \bibinfo{author}{S.~Perkin}, \bibinfo{author}{R.~Atkin}, et~al.,
  \bibinfo{journal}{Chemical communications} \bibinfo{volume}{53}
  (\bibinfo{year}{2017}) \bibinfo{pages}{1214--1224}.
\bibitem[{Tournassat et~al.(2016)Tournassat, Bourg, Holmboe, Sposito, and
  Steefel}]{tournassat2016molecular}
\bibinfo{author}{C.~Tournassat}, \bibinfo{author}{I.~C. Bourg},
  \bibinfo{author}{M.~Holmboe}, \bibinfo{author}{G.~Sposito},
  \bibinfo{author}{C.~I. Steefel}, \bibinfo{journal}{Clays and Clay Minerals}
  \bibinfo{volume}{64} (\bibinfo{year}{2016}) \bibinfo{pages}{374--388}.
\bibitem[{Somers et~al.(2013)Somers, Howlett, MacFarlane, and
  Forsyth}]{somers2013review}
\bibinfo{author}{A.~E. Somers}, \bibinfo{author}{P.~C. Howlett},
  \bibinfo{author}{D.~R. MacFarlane}, \bibinfo{author}{M.~Forsyth},
  \bibinfo{journal}{Lubricants} \bibinfo{volume}{1} (\bibinfo{year}{2013})
  \bibinfo{pages}{3--21}.
\bibitem[{Lee et~al.(2017)Lee, Perez-Martinez, Smith, and
  Perkin}]{perez2017scaling}
\bibinfo{author}{A.~A. Lee}, \bibinfo{author}{C.~S. Perez-Martinez},
  \bibinfo{author}{A.~M. Smith}, \bibinfo{author}{S.~Perkin},
  \bibinfo{journal}{Physical review letters} \bibinfo{volume}{119}
  (\bibinfo{year}{2017}) \bibinfo{pages}{026002}.
\bibitem[{Han et~al.(2006)Han, Alsayed, Nobili, Zhang, Lubensky, and
  Yodh}]{han2006brownian}
\bibinfo{author}{Y.~Han}, \bibinfo{author}{A.~M. Alsayed},
  \bibinfo{author}{M.~Nobili}, \bibinfo{author}{J.~Zhang},
  \bibinfo{author}{T.~C. Lubensky}, \bibinfo{author}{A.~G. Yodh},
  \bibinfo{journal}{Science} \bibinfo{volume}{314} (\bibinfo{year}{2006})
  \bibinfo{pages}{626--630}.
\bibitem[{Pande and Smith(2015)}]{pande2015forces}
\bibinfo{author}{J.~Pande}, \bibinfo{author}{A.-S. Smith},
  \bibinfo{journal}{Soft Matter} \bibinfo{volume}{11} (\bibinfo{year}{2015})
  \bibinfo{pages}{2364--2371}.
\bibitem[{Mittal et~al.(2008)Mittal, Truskett, Errington, and
  Hummer}]{mittal2008layering}
\bibinfo{author}{J.~Mittal}, \bibinfo{author}{T.~M. Truskett},
  \bibinfo{author}{J.~R. Errington}, \bibinfo{author}{G.~Hummer},
  \bibinfo{journal}{Physical Review Letters} \bibinfo{volume}{100}
  (\bibinfo{year}{2008}) \bibinfo{pages}{145901}.
\bibitem[{Fern{\'a}ndez et~al.(2004)Fern{\'a}ndez, Vrabec, and
  Hasse}]{fernandez2004self}
\bibinfo{author}{G.~Fern{\'a}ndez}, \bibinfo{author}{J.~Vrabec},
  \bibinfo{author}{H.~Hasse}, \bibinfo{journal}{International Journal of
  Thermophysics} \bibinfo{volume}{25} (\bibinfo{year}{2004})
  \bibinfo{pages}{175--186}.
\bibitem[{Nordanger et~al.(2022)Nordanger, Morozov, and
  Stenhammar}]{nordanger2022anisotropic}
\bibinfo{author}{H.~Nordanger}, \bibinfo{author}{A.~Morozov},
  \bibinfo{author}{J.~Stenhammar}, \bibinfo{journal}{Physical Review Fluids}
  \bibinfo{volume}{7} (\bibinfo{year}{2022}) \bibinfo{pages}{013103}.
\bibitem[{Smith et~al.(2016)Smith, Lee, and Perkin}]{smith2016electrostatic}
\bibinfo{author}{A.~M. Smith}, \bibinfo{author}{A.~A. Lee},
  \bibinfo{author}{S.~Perkin}, \bibinfo{journal}{The journal of physical
  chemistry letters} \bibinfo{volume}{7} (\bibinfo{year}{2016})
  \bibinfo{pages}{2157--2163}.
\bibitem[{Fedorov and Kornyshev(2008)}]{fedorov2008towards}
\bibinfo{author}{M.~V. Fedorov}, \bibinfo{author}{A.~A. Kornyshev},
  \bibinfo{journal}{Electrochimica Acta} \bibinfo{volume}{53}
  (\bibinfo{year}{2008}) \bibinfo{pages}{6835--6840}.
\bibitem[{Merlet et~al.(2014)Merlet, Limmer, Salanne, Van~Roij, Madden,
  Chandler, and Rotenberg}]{merlet2014electric}
\bibinfo{author}{C.~Merlet}, \bibinfo{author}{D.~T. Limmer},
  \bibinfo{author}{M.~Salanne}, \bibinfo{author}{R.~Van~Roij},
  \bibinfo{author}{P.~A. Madden}, \bibinfo{author}{D.~Chandler},
  \bibinfo{author}{B.~Rotenberg}, \bibinfo{journal}{The Journal of Physical
  Chemistry C} \bibinfo{volume}{118} (\bibinfo{year}{2014})
  \bibinfo{pages}{18291--18298}.
\bibitem[{Salanne and Madden(2011)}]{salanne2011polarization}
\bibinfo{author}{M.~Salanne}, \bibinfo{author}{P.~A. Madden},
  \bibinfo{journal}{Molecular Physics} \bibinfo{volume}{109}
  (\bibinfo{year}{2011}) \bibinfo{pages}{2299--2315}.
\bibitem[{Kondrat et~al.(2014)Kondrat, Wu, Qiao, and
  Kornyshev}]{kondrat2014accelerating}
\bibinfo{author}{S.~Kondrat}, \bibinfo{author}{P.~Wu},
  \bibinfo{author}{R.~Qiao}, \bibinfo{author}{A.~A. Kornyshev},
  \bibinfo{journal}{Nature materials} \bibinfo{volume}{13}
  (\bibinfo{year}{2014}) \bibinfo{pages}{387--393}.
\bibitem[{Salanne et~al.(2012)Salanne, Rotenberg, Jahn, Vuilleumier, Simon, and
  Madden}]{salanne2012including}
\bibinfo{author}{M.~Salanne}, \bibinfo{author}{B.~Rotenberg},
  \bibinfo{author}{S.~Jahn}, \bibinfo{author}{R.~Vuilleumier},
  \bibinfo{author}{C.~Simon}, \bibinfo{author}{P.~A. Madden},
  \bibinfo{journal}{Theoretical Chemistry Accounts} \bibinfo{volume}{131}
  (\bibinfo{year}{2012}) \bibinfo{pages}{1--16}.
\bibitem[{Salanne(2015)}]{salanne2015simulations}
\bibinfo{author}{M.~Salanne}, \bibinfo{journal}{Physical Chemistry Chemical
  Physics} \bibinfo{volume}{17} (\bibinfo{year}{2015})
  \bibinfo{pages}{14270--14279}.
\bibitem[{P{\'a}dua et~al.(2007)P{\'a}dua, Costa~Gomes, and
  Canongia~Lopes}]{padua2007molecular}
\bibinfo{author}{A.~A. P{\'a}dua}, \bibinfo{author}{M.~F. Costa~Gomes},
  \bibinfo{author}{J.~N. Canongia~Lopes}, \bibinfo{journal}{Accounts of
  Chemical Research} \bibinfo{volume}{40} (\bibinfo{year}{2007})
  \bibinfo{pages}{1087--1096}.
\bibitem[{Canongia~Lopes and P{\'a}dua(2006)}]{canongia2006nanostructural}
\bibinfo{author}{J.~N. Canongia~Lopes}, \bibinfo{author}{A.~A. P{\'a}dua},
  \bibinfo{journal}{The Journal of Physical Chemistry B} \bibinfo{volume}{110}
  (\bibinfo{year}{2006}) \bibinfo{pages}{3330--3335}.
\bibitem[{Canongia~Lopes et~al.(2004)Canongia~Lopes, Deschamps, and
  P{\'a}dua}]{canongia2004modeling}
\bibinfo{author}{J.~N. Canongia~Lopes}, \bibinfo{author}{J.~Deschamps},
  \bibinfo{author}{A.~A. P{\'a}dua}, \bibinfo{journal}{The Journal of Physical
  Chemistry B} \bibinfo{volume}{108} (\bibinfo{year}{2004})
  \bibinfo{pages}{2038--2047}.
\bibitem[{Canongia~Lopes and P{\'a}dua(2004)}]{canongia2004molecular}
\bibinfo{author}{J.~N. Canongia~Lopes}, \bibinfo{author}{A.~A. P{\'a}dua},
  \bibinfo{journal}{The Journal of Physical Chemistry B} \bibinfo{volume}{108}
  (\bibinfo{year}{2004}) \bibinfo{pages}{16893--16898}.
\bibitem[{Horstmann et~al.(2022)Horstmann, Hecht, Kloth, and
  Vogel}]{horstmann2022structural}
\bibinfo{author}{R.~Horstmann}, \bibinfo{author}{L.~Hecht},
  \bibinfo{author}{S.~Kloth}, \bibinfo{author}{M.~Vogel},
  \bibinfo{journal}{Langmuir}  (\bibinfo{year}{2022}).
\bibitem[{Kubo(1957)}]{kubo1957statistical}
\bibinfo{author}{R.~Kubo}, \bibinfo{journal}{Journal of the Physical Society of
  Japan} \bibinfo{volume}{12} (\bibinfo{year}{1957}) \bibinfo{pages}{570--586}.
\bibitem[{Green(1954)}]{green1954markoff}
\bibinfo{author}{M.~S. Green}, \bibinfo{journal}{The Journal of Chemical
  Physics} \bibinfo{volume}{22} (\bibinfo{year}{1954})
  \bibinfo{pages}{398--413}.
\bibitem[{Zhou and Miller(1996)}]{zhou1996green}
\bibinfo{author}{Y.~Zhou}, \bibinfo{author}{G.~H. Miller},
  \bibinfo{journal}{The Journal of Physical Chemistry} \bibinfo{volume}{100}
  (\bibinfo{year}{1996}) \bibinfo{pages}{5516--5524}.
\bibitem[{Fong et~al.(2021)Fong, Self, McCloskey, and Persson}]{fong2021ion}
\bibinfo{author}{K.~D. Fong}, \bibinfo{author}{J.~Self}, \bibinfo{author}{B.~D.
  McCloskey}, \bibinfo{author}{K.~A. Persson},
  \bibinfo{journal}{Macromolecules} \bibinfo{volume}{54} (\bibinfo{year}{2021})
  \bibinfo{pages}{2575--2591}.
\bibitem[{Zwanzig and Bixon(1970)}]{zwanzig1970hydrodynamic}
\bibinfo{author}{R.~Zwanzig}, \bibinfo{author}{M.~Bixon},
  \bibinfo{journal}{Physical Review A} \bibinfo{volume}{2}
  (\bibinfo{year}{1970}) \bibinfo{pages}{2005}.
\bibitem[{Micheletti et~al.(2008)Micheletti, Bussi, and
  Laio}]{micheletti2008optimal}
\bibinfo{author}{C.~Micheletti}, \bibinfo{author}{G.~Bussi},
  \bibinfo{author}{A.~Laio}, \bibinfo{journal}{The Journal of Chemical Physics}
  \bibinfo{volume}{129} (\bibinfo{year}{2008}) \bibinfo{pages}{074105}.
\bibitem[{Vella(2019)}]{vella2019fick}
\bibinfo{author}{J.~R. Vella}, \bibinfo{journal}{Journal of Chemical \&
  Engineering Data} \bibinfo{volume}{64} (\bibinfo{year}{2019})
  \bibinfo{pages}{3672--3681}.
\bibitem[{Sicard et~al.(2021)Sicard, Koskin, Annibale, and
  Rosta}]{sicard2021position}
\bibinfo{author}{F.~Sicard}, \bibinfo{author}{V.~Koskin},
  \bibinfo{author}{A.~Annibale}, \bibinfo{author}{E.~Rosta},
  \bibinfo{journal}{Journal of Chemical Theory and Computation}
  \bibinfo{volume}{17} (\bibinfo{year}{2021}) \bibinfo{pages}{2022--2033}.
\bibitem[{Uhlenbeck and Ornstein(1930)}]{uhlenbeck1930theory}
\bibinfo{author}{G.~E. Uhlenbeck}, \bibinfo{author}{L.~S. Ornstein},
  \bibinfo{journal}{Physical Review} \bibinfo{volume}{36}
  (\bibinfo{year}{1930}) \bibinfo{pages}{823}.
\bibitem[{T{\"u}rkcan et~al.(2012)T{\"u}rkcan, Alexandrou, and
  Masson}]{turkcan2012bayesian}
\bibinfo{author}{S.~T{\"u}rkcan}, \bibinfo{author}{A.~Alexandrou},
  \bibinfo{author}{J.-B. Masson}, \bibinfo{journal}{Biophysical Journal}
  \bibinfo{volume}{102} (\bibinfo{year}{2012}) \bibinfo{pages}{2288--2298}.
\bibitem[{Nagai et~al.(2020)Nagai, Tsurumaki, Urano, Fujimoto, Shinoda, and
  Okazaki}]{nagai2020position}
\bibinfo{author}{T.~Nagai}, \bibinfo{author}{S.~Tsurumaki},
  \bibinfo{author}{R.~Urano}, \bibinfo{author}{K.~Fujimoto},
  \bibinfo{author}{W.~Shinoda}, \bibinfo{author}{S.~Okazaki},
  \bibinfo{journal}{Journal of Chemical Theory and Computation}
  \bibinfo{volume}{16} (\bibinfo{year}{2020}) \bibinfo{pages}{7239--7254}.
\bibitem[{Holmboe and Bourg(2014)}]{holmboe2014molecular}
\bibinfo{author}{M.~Holmboe}, \bibinfo{author}{I.~C. Bourg},
  \bibinfo{journal}{The Journal of Physical Chemistry C} \bibinfo{volume}{118}
  (\bibinfo{year}{2014}) \bibinfo{pages}{1001--1013}.
\bibitem[{Hinczewski et~al.(2010)Hinczewski, von Hansen, Dzubiella, and
  Netz}]{hinczewski2010diffusivity}
\bibinfo{author}{M.~Hinczewski}, \bibinfo{author}{Y.~von Hansen},
  \bibinfo{author}{J.~Dzubiella}, \bibinfo{author}{R.~R. Netz},
  \bibinfo{journal}{The Journal of Chemical Physics} \bibinfo{volume}{132}
  (\bibinfo{year}{2010}) \bibinfo{pages}{06B615}.
\bibitem[{Shalchi(2011)}]{shalchi2011applicability}
\bibinfo{author}{A.~Shalchi}, \bibinfo{journal}{Physical Review E}
  \bibinfo{volume}{83} (\bibinfo{year}{2011}) \bibinfo{pages}{046402}.
\bibitem[{Comer et~al.(2013)Comer, Chipot, and
  Gonz{\'a}lez-Nilo}]{comer2013calculating}
\bibinfo{author}{J.~Comer}, \bibinfo{author}{C.~Chipot}, \bibinfo{author}{F.~D.
  Gonz{\'a}lez-Nilo}, \bibinfo{journal}{Journal of Chemical Theory and
  Computation} \bibinfo{volume}{9} (\bibinfo{year}{2013})
  \bibinfo{pages}{876--882}.
\bibitem[{Best and Hummer(2010)}]{best2010coordinate}
\bibinfo{author}{R.~B. Best}, \bibinfo{author}{G.~Hummer},
  \bibinfo{journal}{Proceedings of the National Academy of Sciences}
  \bibinfo{volume}{107} (\bibinfo{year}{2010}) \bibinfo{pages}{1088--1093}.
\bibitem[{Bourg and Sposito(2011)}]{bourg2011molecular}
\bibinfo{author}{I.~C. Bourg}, \bibinfo{author}{G.~Sposito},
  \bibinfo{journal}{Journal of Colloid and Interface Science}
  \bibinfo{volume}{360} (\bibinfo{year}{2011}) \bibinfo{pages}{701--715}.
\bibitem[{Buchete and Hummer(2008)}]{buchete2008coarse}
\bibinfo{author}{N.-V. Buchete}, \bibinfo{author}{G.~Hummer},
  \bibinfo{journal}{The Journal of Physical Chemistry B} \bibinfo{volume}{112}
  (\bibinfo{year}{2008}) \bibinfo{pages}{6057--6069}.
\bibitem[{Bourg and Steefel(2012)}]{bourg2012molecular}
\bibinfo{author}{I.~C. Bourg}, \bibinfo{author}{C.~I. Steefel},
  \bibinfo{journal}{The Journal of Physical Chemistry C} \bibinfo{volume}{116}
  (\bibinfo{year}{2012}) \bibinfo{pages}{11556--11564}.
\bibitem[{Rosta and Hummer(2015)}]{rosta2015free}
\bibinfo{author}{E.~Rosta}, \bibinfo{author}{G.~Hummer},
  \bibinfo{journal}{Journal of Chemical Theory and Computation}
  \bibinfo{volume}{11} (\bibinfo{year}{2015}) \bibinfo{pages}{276--285}.
\bibitem[{Husic and Pande(2018)}]{husic2018markov}
\bibinfo{author}{B.~E. Husic}, \bibinfo{author}{V.~S. Pande},
  \bibinfo{journal}{Journal of the American Chemical Society}
  \bibinfo{volume}{140} (\bibinfo{year}{2018}) \bibinfo{pages}{2386--2396}.
\bibitem[{Renkin(1954)}]{renkin1954filtration}
\bibinfo{author}{E.~M. Renkin}, \bibinfo{journal}{The Journal of General
  Physiology} \bibinfo{volume}{38} (\bibinfo{year}{1954}) \bibinfo{pages}{225}.
\bibitem[{Hummer(2005)}]{hummer2005position}
\bibinfo{author}{G.~Hummer}, \bibinfo{journal}{New Journal of Physics}
  \bibinfo{volume}{7} (\bibinfo{year}{2005}) \bibinfo{pages}{34}.
\bibitem[{Rohrdanz et~al.(2011)Rohrdanz, Zheng, Maggioni, and
  Clementi}]{rohrdanz2011determination}
\bibinfo{author}{M.~A. Rohrdanz}, \bibinfo{author}{W.~Zheng},
  \bibinfo{author}{M.~Maggioni}, \bibinfo{author}{C.~Clementi},
  \bibinfo{journal}{The Journal of Chemical Physics} \bibinfo{volume}{134}
  (\bibinfo{year}{2011}) \bibinfo{pages}{03B624}.
\bibitem[{Pérez-Hernández et~al.(2013)Pérez-Hernández, Paul, Giorgino,
  De~Fabritiis, and Noé}]{perez2013slowparameters}
\bibinfo{author}{G.~Pérez-Hernández}, \bibinfo{author}{F.~Paul},
  \bibinfo{author}{T.~Giorgino}, \bibinfo{author}{G.~De~Fabritiis},
  \bibinfo{author}{F.~Noé}, \bibinfo{journal}{The Journal of Chemical Physics}
  \bibinfo{volume}{139} (\bibinfo{year}{2013}) \bibinfo{pages}{015102}.
  \DOIprefix\doi{10.1063/1.4811489}.
\bibitem[{Voisinne et~al.(2010)Voisinne, Alexandrou, and
  Masson}]{voisinne2010quantifying}
\bibinfo{author}{G.~Voisinne}, \bibinfo{author}{A.~Alexandrou},
  \bibinfo{author}{J.-B. Masson}, \bibinfo{journal}{Biophysical Journal}
  \bibinfo{volume}{98} (\bibinfo{year}{2010}) \bibinfo{pages}{596--605}.
\bibitem[{Bicout and Szabo(1998)}]{bicout1998electron}
\bibinfo{author}{D.~Bicout}, \bibinfo{author}{A.~Szabo}, \bibinfo{journal}{The
  Journal of Chemical Physics} \bibinfo{volume}{109} (\bibinfo{year}{1998})
  \bibinfo{pages}{2325--2338}.
\bibitem[{Vu{\v{c}}emilovi{\'c}-Alagi{\'c}
  et~al.(2019)Vu{\v{c}}emilovi{\'c}-Alagi{\'c}, Banhatti, Stepi{\'c}, Wick,
  Berger, Gaimann, Baer, Harting, Smith, and Smith}]{vucemilovic2019insights}
\bibinfo{author}{N.~Vu{\v{c}}emilovi{\'c}-Alagi{\'c}}, \bibinfo{author}{R.~D.
  Banhatti}, \bibinfo{author}{R.~Stepi{\'c}}, \bibinfo{author}{C.~R. Wick},
  \bibinfo{author}{D.~Berger}, \bibinfo{author}{M.~U. Gaimann},
  \bibinfo{author}{A.~Baer}, \bibinfo{author}{J.~Harting},
  \bibinfo{author}{D.~M. Smith}, \bibinfo{author}{A.-S. Smith},
  \bibinfo{journal}{Journal of Colloid and Interface Science}
  \bibinfo{volume}{553} (\bibinfo{year}{2019}) \bibinfo{pages}{350--363}.
\bibitem[{Vu{\v{c}}emilovi{\'c}-Alagi{\'c}(2021)}]{vuvcemilovic2021computational}
\bibinfo{author}{N.~Vu{\v{c}}emilovi{\'c}-Alagi{\'c}},
  \bibinfo{title}{Computational study of the physical and interfacial
  properties of Imidazolium-based ionic liquids}, Ph.D. thesis,
  Friedrich-Alexander-Universitaet Erlangen-Nuernberg (Germany),
  \bibinfo{year}{2021}.
\bibitem[{Belousov et~al.(2022)Belousov, Hassanali, and
  Rold{\'a}n}]{belousov2022statistical}
\bibinfo{author}{R.~Belousov}, \bibinfo{author}{A.~Hassanali},
  \bibinfo{author}{{\'E}.~Rold{\'a}n}, \bibinfo{journal}{Physical Review E}
  \bibinfo{volume}{106} (\bibinfo{year}{2022}) \bibinfo{pages}{014103}.
\bibitem[{Atkin and Warr(2007)}]{atkin2007structure}
\bibinfo{author}{R.~Atkin}, \bibinfo{author}{G.~G. Warr}, \bibinfo{journal}{The
  Journal of Physical Chemistry C} \bibinfo{volume}{111} (\bibinfo{year}{2007})
  \bibinfo{pages}{5162--5168}.
\bibitem[{Brklja{\v{c}}a et~al.(2015)Brklja{\v{c}}a, Klimczak, Milicevic,
  Weisser, Taccardi, Wasserscheid, Smith, Magerl, and
  Smith}]{brkljaca2015complementary}
\bibinfo{author}{Z.~Brklja{\v{c}}a}, \bibinfo{author}{M.~Klimczak},
  \bibinfo{author}{Z.~Milicevic}, \bibinfo{author}{M.~Weisser},
  \bibinfo{author}{N.~Taccardi}, \bibinfo{author}{P.~Wasserscheid},
  \bibinfo{author}{D.~M. Smith}, \bibinfo{author}{A.~Magerl},
  \bibinfo{author}{A.-S. Smith}, \bibinfo{journal}{The Journal of Physical
  Chemistry Letters} \bibinfo{volume}{6} (\bibinfo{year}{2015})
  \bibinfo{pages}{549--555}.
\bibitem[{Vucemilovic-Alagic et~al.(2019)Vucemilovic-Alagic, Banhatti, Stepic,
  Wick, Berger, Gaimann, Bear, Harting, Smith, and
  Smith}]{vucemilovic2019structural}
\bibinfo{author}{N.~Vucemilovic-Alagic}, \bibinfo{author}{R.~D. Banhatti},
  \bibinfo{author}{R.~Stepic}, \bibinfo{author}{C.~R. Wick},
  \bibinfo{author}{D.~Berger}, \bibinfo{author}{M.~Gaimann},
  \bibinfo{author}{A.~Bear}, \bibinfo{author}{J.~Harting},
  \bibinfo{author}{D.~M. Smith}, \bibinfo{author}{A.-S. Smith},
  \bibinfo{journal}{arXiv preprint arXiv:1905.06009}  (\bibinfo{year}{2019}).
\bibitem[{D{\"u}nweg and Kremer(1993)}]{dunweg1993molecular}
\bibinfo{author}{B.~D{\"u}nweg}, \bibinfo{author}{K.~Kremer},
  \bibinfo{journal}{The Journal of Chemical Physics} \bibinfo{volume}{99}
  (\bibinfo{year}{1993}) \bibinfo{pages}{6983--6997}.
\bibitem[{Yeh and Hummer(2004)}]{yeh2004system}
\bibinfo{author}{I.-C. Yeh}, \bibinfo{author}{G.~Hummer}, \bibinfo{journal}{The
  Journal of Physical Chemistry B} \bibinfo{volume}{108} (\bibinfo{year}{2004})
  \bibinfo{pages}{15873--15879}.
\bibitem[{Prakash et~al.(2017)Prakash, Lemaire, Caruel, Lewerenz, de~Leeuw,
  Di~Tommaso, and Naili}]{prakash2017anisotropic}
\bibinfo{author}{M.~Prakash}, \bibinfo{author}{T.~Lemaire},
  \bibinfo{author}{M.~Caruel}, \bibinfo{author}{M.~Lewerenz},
  \bibinfo{author}{N.~H. de~Leeuw}, \bibinfo{author}{D.~Di~Tommaso},
  \bibinfo{author}{S.~Naili}, \bibinfo{journal}{Physics and Chemistry of
  Minerals} \bibinfo{volume}{44} (\bibinfo{year}{2017})
  \bibinfo{pages}{509--519}.
\bibitem[{Baer et~al.(2022)Baer, Malgaretti, Kaspereit, Harting, and
  Smith}]{baer2022modelling}
\bibinfo{author}{A.~Baer}, \bibinfo{author}{P.~Malgaretti},
  \bibinfo{author}{M.~Kaspereit}, \bibinfo{author}{J.~Harting},
  \bibinfo{author}{A.-S. Smith}, \bibinfo{journal}{Journal of Molecular
  Liquids}  (\bibinfo{year}{2022}) \bibinfo{pages}{120636}.
\bibitem[{Gentile et~al.(2015)Gentile, De~Santo, D’Avino, Rossi, Romeo,
  Greco, Netti, and Maffettone}]{gentile2015hindered}
\bibinfo{author}{F.~S. Gentile}, \bibinfo{author}{I.~De~Santo},
  \bibinfo{author}{G.~D’Avino}, \bibinfo{author}{L.~Rossi},
  \bibinfo{author}{G.~Romeo}, \bibinfo{author}{F.~Greco},
  \bibinfo{author}{P.~A. Netti}, \bibinfo{author}{P.~L. Maffettone},
  \bibinfo{journal}{Journal of colloid and interface science}
  \bibinfo{volume}{447} (\bibinfo{year}{2015}) \bibinfo{pages}{25--32}.
\bibitem[{Feitosa and Mesquita(1991)}]{feitosa1991wall}
\bibinfo{author}{M.~I.~M. Feitosa}, \bibinfo{author}{O.~N. Mesquita},
  \bibinfo{journal}{Physical Review A} \bibinfo{volume}{44}
  (\bibinfo{year}{1991}) \bibinfo{pages}{6677}.
\bibitem[{Tsimpanogiannis et~al.(2019)Tsimpanogiannis, Moultos, Franco, Spera,
  Erd{\H{o}}s, and Economou}]{tsimpanogiannis2019self}
\bibinfo{author}{I.~N. Tsimpanogiannis}, \bibinfo{author}{O.~A. Moultos},
  \bibinfo{author}{L.~F. Franco}, \bibinfo{author}{M.~B. d.~M. Spera},
  \bibinfo{author}{M.~Erd{\H{o}}s}, \bibinfo{author}{I.~G. Economou},
  \bibinfo{journal}{Molecular Simulation} \bibinfo{volume}{45}
  (\bibinfo{year}{2019}) \bibinfo{pages}{425--453}.
\bibitem[{Bevan and Prieve(2000)}]{bevan2000hindered}
\bibinfo{author}{M.~A. Bevan}, \bibinfo{author}{D.~C. Prieve},
  \bibinfo{journal}{The Journal of Chemical Physics} \bibinfo{volume}{113}
  (\bibinfo{year}{2000}) \bibinfo{pages}{1228--1236}.
\bibitem[{Goldman et~al.(1967)Goldman, Cox, and Brenner}]{goldman1967slow}
\bibinfo{author}{A.~J. Goldman}, \bibinfo{author}{R.~G. Cox},
  \bibinfo{author}{H.~Brenner}, \bibinfo{journal}{Chemical engineering science}
  \bibinfo{volume}{22} (\bibinfo{year}{1967}) \bibinfo{pages}{637--651}.
\bibitem[{Brenner(1961)}]{brenner1961slow}
\bibinfo{author}{H.~Brenner}, \bibinfo{journal}{Chemical engineering science}
  \bibinfo{volume}{16} (\bibinfo{year}{1961}) \bibinfo{pages}{242--251}.
\bibitem[{Cvitkovi{\'c} et~al.(2022)Cvitkovi{\'c}, Ghanti, Raake, and
  Smith}]{cvitkovic2022crowding}
\bibinfo{author}{M.~Cvitkovi{\'c}}, \bibinfo{author}{D.~Ghanti},
  \bibinfo{author}{N.~Raake}, \bibinfo{author}{A.-S. Smith},
  \bibinfo{journal}{The European Physical Journal Plus} \bibinfo{volume}{137}
  (\bibinfo{year}{2022}) \bibinfo{pages}{355}.
\bibitem[{H{\"o}llring et~al.(2023)H{\"o}llring, Baer,
  Vu\v{c}emilovi\'c-Alagi\'c, Smith, and Smith}]{epm_pub}
\bibinfo{author}{K.~H{\"o}llring}, \bibinfo{author}{A.~Baer},
  \bibinfo{author}{N.~Vu\v{c}emilovi\'c-Alagi\'c}, \bibinfo{author}{D.~M.
  Smith}, \bibinfo{author}{A.-S. Smith}, \bibinfo{journal}{Journal of Colloid
  and Interface Science} \bibinfo{volume}{???} (\bibinfo{year}{2023})
  \bibinfo{pages}{???}
\bibitem[{Berg(1993)}]{berg1993random}
\bibinfo{author}{H.~C. Berg}, \bibinfo{title}{Random walks in biology},
  \bibinfo{publisher}{Princeton University Press}, \bibinfo{year}{1993}.
\bibitem[{Bicout et~al.(1998)Bicout, Berezhkovskii, and
  Weiss}]{bicout1998turnover}
\bibinfo{author}{D.~Bicout}, \bibinfo{author}{A.~Berezhkovskii},
  \bibinfo{author}{G.~Weiss}, \bibinfo{journal}{Physica A: Statistical
  Mechanics and its Applications} \bibinfo{volume}{256} (\bibinfo{year}{1998})
  \bibinfo{pages}{342--350}.
\bibitem[{van Hijkoop et~al.(2007)van Hijkoop, Dammers, Malek, and
  Coppens}]{van2007water}
\bibinfo{author}{V.~J. van Hijkoop}, \bibinfo{author}{A.~J. Dammers},
  \bibinfo{author}{K.~Malek}, \bibinfo{author}{M.-O. Coppens},
  \bibinfo{journal}{The Journal of chemical physics} \bibinfo{volume}{127}
  (\bibinfo{year}{2007}) \bibinfo{pages}{08B613}.
\bibitem[{Mercier~Franco et~al.(2016)Mercier~Franco, Castier, and
  Economou}]{mercier2016diffusion}
\bibinfo{author}{L.~F. Mercier~Franco}, \bibinfo{author}{M.~Castier},
  \bibinfo{author}{I.~G. Economou}, \bibinfo{journal}{Journal of Chemical
  Theory and Computation} \bibinfo{volume}{12} (\bibinfo{year}{2016})
  \bibinfo{pages}{5247--5255}.
\bibitem[{Bussi et~al.(2007)Bussi, Donadio, and Parrinello}]{Bussi2007}
\bibinfo{author}{G.~Bussi}, \bibinfo{author}{D.~Donadio},
  \bibinfo{author}{M.~Parrinello}, \bibinfo{journal}{The Journal of chemical
  physics} \bibinfo{volume}{126} (\bibinfo{year}{2007})
  \bibinfo{pages}{014101}.
\bibitem[{Höllring et~al.(2022)Höllring, Baer, Vučemilović-Alagić, Smith,
  and Smith}]{hollring_kevin_2022_7446071}
\bibinfo{author}{K.~Höllring}, \bibinfo{author}{A.~Baer},
  \bibinfo{author}{N.~Vučemilović-Alagić}, \bibinfo{author}{D.~M. Smith},
  \bibinfo{author}{A.-S. Smith}, \bibinfo{title}{{Extracting diffusion profiles
  of particles in slit pores}}, \bibinfo{year}{2022}. \URLprefix
  \url{https://doi.org/10.5281/zenodo.7446071}.
  \DOIprefix\doi{10.5281/zenodo.7446071}.

\end{thebibliography}



\end{document}